\documentclass[aps,pra,superscriptaddress,twocolumn,preprintnumbers,10pt]{revtex4-2}
\usepackage{amssymb}
\usepackage{amsmath}
\usepackage{graphicx}
\usepackage{natbib}
\usepackage{bm}
\usepackage{color}
\usepackage{csquotes}
\usepackage{appendix}
\usepackage{amsfonts}
\usepackage{soul}
\definecolor{blue}{rgb}{0.00,0.07,1.00}
\begin{document}
\title{Dynamics of Quantum Entanglement Between Photon and Phonon Modes in a Coulomb-coupled  Optomechanical Cavity Magnonic Systems}
\author{Muhib Ullah}
\email{muhibqau@hotmail.com}
\affiliation{Zhejiang University/University of Illinois Urbana-Champaign (ZJU-UIUC) Institute, Zhejiang University, International Campus, Haining, Zhejiang, China}
\author{Muhammad Idrees}
\affiliation{Institute of Nonlinear Physics and Department of
Physics, Zhejiang Normal University, Jinhua 321004,
China}
\affiliation{Zhejiang Institute of Photoelectronics and
Zhejiang Institute for Advanced Light Source, Zhejiang Normal
University, Jinhua 321004, China}
\author{Said Mikki}
\email{smikki@illinois.edu}
\affiliation{Zhejiang University/University of Illinois Urbana-Champaign
(ZJU-UIUC) Institute, Zhejiang University, International Campus,
Haining, Zhejiang, China}
\affiliation{Electrical and Computer Engineering Department, University of Illinois Urbana-Champaign, Urbana, Illinois, United States}
\begin{abstract}
Quantum entanglement is a fundamental phenomenon in quantum information science and a crucial resource for advancing quantum technologies, including precision sensing, ultra-secure communication, and quantum computation. In hybrid systems combining cavity optomechanics and magnonics, entanglement between cavity photons, magnons, and mechanical resonators offers unique opportunities for manipulating and controlling the quantum states across different modes via radiation pressure-based light-matter interaction.
In this study, we propose a hybrid optomechanical cavity-magnonic setup that integrates a Yttrium iron garnet (YIG) sphere to mediate photon-magnon coupling via magnetic dipole interaction. The system also incorporates optomechanical and electrostatic interactions to facilitate entanglement between cavity photons and mechanical resonators. A primary focus of this work is to investigate how the magnon’s nonlinear Kerr effect, along with varying optomechanical and electrostatic coupling strengths, influences the entanglement dynamics within the setup.
By analyzing the interaction of nonlinear effects with other coupling parameters, we aim to identify the optimal conditions for both initiating and sustaining robust entanglement between the motional modes (phonons) of two Coulomb-coupled mechanical resonators, as well as between the cavity photon and a mechanical resonator. Our computational model predicts that incorporating Coulomb interactions and magnon coupling into the setup significantly enhances the degree and tunability of the attained entanglement, making it more resilient to external thermal baths at around 3K.
This study addresses fundamental questions regarding the conditions and techniques needed to optimize and sustain phonon-phonon entanglement in complex hybrid quantum systems. The findings provide valuable insights that could be useful for the development of quantum memories in continuous-variable quantum information processing and other quantum-enhanced technologies.
\end{abstract}
\maketitle
\section{Introduction}
Quantum entanglement \cite{horodecki2009quantum}, a fundamental resource in quantum information processing, has attracted significant attention in quantum physics. Over the past few decades, various physical platforms, including trapped ions, atomic ensembles, and photonic systems \cite{yang2005efficient,deb2011quantum,mirkhalaf2017entanglement,bittencourt2016quantum,osawa2021controllable}, have been effectively utilized for the generation and manipulation of entangled states. These developments are particularly relevant to a broad range of quantum information applications, such as quantum communication and teleportation \cite{bennett1993teleporting,boschi1998experimental}, quantum computing \cite{divincenzo1995quantum}, quantum dense coding \cite{braunstein2000dense}, and quantum cryptography \cite{bennett1992quantum,fuchs1997optimal}.
In recent years, a versatile framework within cavity optomechanics has emerged for engineering interactions between diverse quantum systems, creating opportunities for the development of hybrid quantum networks that leverage the unique properties of each system \cite{aspelmeyer2012quantum,aspelmeyer2014cavity,bochmann2013nanomechanical,andrews2014bidirectional,balram2016coherent}. These networks enable coherent quantum communication across distinct platforms \cite{stannigel2010optomechanical,dong2015optomechanical}. Various force-mediated coupling mechanisms, such as magnetostrictive interactions \cite{zhang2016cavity,colombano2020ferromagnetic}, radiation pressure \cite{shen2016experimental,shen2018reconfigurable}, electrostatic forces \cite{andrews2014bidirectional,higginbotham2018harnessing,arnold2020converting}, and piezoelectric forces \cite{bochmann2013nanomechanical,jiang2020efficient,forsch2020microwave,han2020cavity}, have been employed to interconnect mechanical motion with magnonic excitations and optical or microwave photonic modes.

Hybrid quantum systems, which integrate two or more distinct subsystems, offer a promising avenue for exploiting the advantages of various quantum platforms. Magnetic materials, for example, possess unique characteristics such as long lifetimes, high spin densities \cite{sun2021remote}, and exceptional tunability \cite{yuan2020enhancement}. Among these, Yttrium iron garnet (YIG) crystals stand out due to their straightforward magnetization process and extremely low dissipation rates \cite{li2019entangling,zhang2019quantum,wang2023simulating}, making them ideal candidates for applications in quantum storage \cite{setodeh2023magnon}, quantum communication \cite{wan2024quantum}, and quantum computing \cite{morris2017strong}. Mechanical oscillators, on the other hand, are highly suited for precision measurement of physical quantities such as displacement and acceleration \cite{li2021cavity}. Thus, integrating mechanical oscillators with YIG crystals in a hybrid system offers significant potential for enabling quantum transduction between photons and phonons \cite{yang2020nonreciprocal}.

Cavity magnonics \cite{li2019entangling,li2018magnon,li2019squeezed,kong2019magnon,ullah2024nonreciprocal,zhang2019quantum,yuan2017magnon}, a now well-established interdisciplinary research field that investigates the interaction between magnons and electromagnetic fields within a resonant cavity, has garnered considerable attention due to its potential as a platform for exploring quantum correlations. The distinct advantages of magnons, such as their long lifetimes and high tunability, make them particularly attractive in this context. Numerous theoretical models \cite{soykal2010strong,soykal2010size,zare2015magnetic} and experimental setups \cite{goryachev2014high,huebl2013high,zhang2014strongly} have been developed to study cavity magnonic systems. In particular, a strong coupling has been demonstrated between the microwave cavity mode and the Kittel mode \cite{kittel1948theory}, which is a spatially uniform mode of spin waves in a small YIG sample characterized by a low damping rate—a significant achievement considering the challenges typically faced by spin ensembles in paramagnetic materials. This cavity magnonic setup exhibits a variety of intriguing phenomena, including cavity magnon polaritons \cite{huebl2013high,zhang2014strongly,kong2019magnetically}, the magnon Kerr effect \cite{wang2016magnon,wang2018bistability,liu2018magnon}, non-classical states \cite{li2018magnon,li2019entangling,li2019squeezed,yang2021controlling}, magnon-induced transparency \cite{wang2018magnon}, optical diodes \cite{kong2019magnon}, bidirectional microwave-optical conversion \cite{hisatomi2016bidirectional}, and cooperative polariton dynamics \cite{yao2017cooperative}.

Quantum entanglement between \textit{motional} quantum states serves as an effective platform for exploring the boundaries between classical and quantum physics \cite{pinard2005entangling,ge2013entanglement}, providing a unique opportunity to investigate quantum behaviors and coherence at scales previously thought to be governed solely by classical mechanics. Numerous proposals advocate for the use of optomechanical cavity systems to entangle the motional quantum states of two massive mechanical oscillators \cite{mancini2002entangling,pinard2005entangling,tan2013dissipation,woolley2014two,li2015generation}. In these configurations, two movable mirrors are integrated into a resonant optical cavity, where the radiation pressure forces within the cavity can be manipulated to create strong correlations and even entanglement between the motions of the mirrors. Notably, this method eliminates the need for direct interaction between the oscillators, enabling them to remain spatially separated while still achieving entanglement. 
In addition to the nonlinearity introduced by the optomechanical interaction in an optomechanical system, the Coulomb interaction represents another avenue for influencing the system's nonlinearity and enhancing entanglement between mechanical resonators \cite{bai2021generation,sohail2020enhancement}. Furthermore, Pan et al. \cite{pan2023enhanced} investigated the generation of entanglement within a hybrid optomechanical system that integrates both an optical parametric amplifier and Coulomb interactions, demonstrating how these combined elements can enhance quantum correlations among the system's components. Similarly, Mekonnen \textit{et al.} have theoretically explored the quantum correlations in a parametric amplifier-assisted optomechanical system having Coulomb-type interaction \cite{mekonnen2023quantum}.

Based on the preceding discussion, an intriguing avenue that warrants further exploration is the role of a Coulomb-enabled optomechanical cavity-magnonic system in influencing and enhancing entanglement between two motional quantum states. We propose a model of an optomechanical cavity system incorporating a YIG sphere positioned within it. This setup features two perfectly reflecting mechanical resonators that are oppositely charged via a bias voltage, thereby establishing electrostatic/Coulomb coupling between them. 
Generally, microwave fixed cavity setups are employed in scenarios involving enclosed YIG spheres, where cavity photons couple with magnons through magnetic dipole interactions \cite{li2018magnon}. In addition to the magnetic dipole coupling between photons and magnons, our proposed model facilitates the interaction of microwave cavity photons with the phonons of a mechanical resonator via a radiation pressure \cite{chen2023nonreciprocal}. The YIG sphere is excited by an external magnetic bias field $ B_0 $, generating the appropriate quanta of spin-wave excitations. When the bias field is sufficiently strong, it excites the magnons, making nonlinear interactions between them significant and leading to a self-Kerr effect. This nonlinearity alters the magnon frequency, resulting in a shift in the resonance frequency of the magnonic system. The combined effects of magnetic nonlinearity and the magnon self-Kerr effect significantly modify the system's dynamics, ultimately enhancing the generation of quantum entanglement across various degrees of freedom.

In the linearized quantum Langevin equations (QLEs), the term related to squeezing is negligible compared to the number-dependent frequency shift we have included, especially in the low-magnon-number regime. Including squeezing would require a more complex theoretical framework, but our simplified model effectively captures the main physical effects. We focus on the frequency-shifting effect of magnon self-Kerr nonlinearity, while neglecting the squeezing effect, which is valid in the regime of weak nonlinearities. Although exploring squeezing effects is an interesting avenue for future research, it lies outside the scope of this study.
Electrostatic control serves as the primary means to couple two motional modes, which can subsequently be employed to establish and enhance entanglement between them. Our findings not only facilitate the generation of quantum entanglement among various modes of the nonlinear hybrid system but also enable efficient control and enhancement of this entanglement through practically feasible parameters. This capability is crucial for the development of quantum memories \cite{bozkurt2025mechanical} in quantum information processing and quantum communication.

The paper is organized as follows: In Section \ref{model}, we provide a detailed description of our proposed model and derive the equations of motion from the total Hamiltonian of the dissipative hybrid system using the Heisenberg approach, also known as the quantum Langevin equation. We also apply the Lyapunov equation to the fluctuation equations to obtain the covariance matrix for the modes subjected to entanglement. To quantify and evaluate the entanglement between the two motional modes, we employ the logarithmic negativity relation. In Section \ref{results}, we conduct a comprehensive investigation into the existence of quantum entanglement between the two mechanical resonators and between a cavity photon and a mechanical resonator through various numerical examples. Finally, in Section \ref{outlook}, we present the concluding remarks on our work.

\section{Description of the Setup, Hamiltonian, and the Quantum Langevin Equation Solution}
\label{model}

\subsection{The Setup and its Hamiltonian}
\begin{figure}[t!]
\centering
\includegraphics[width=0.95\columnwidth]{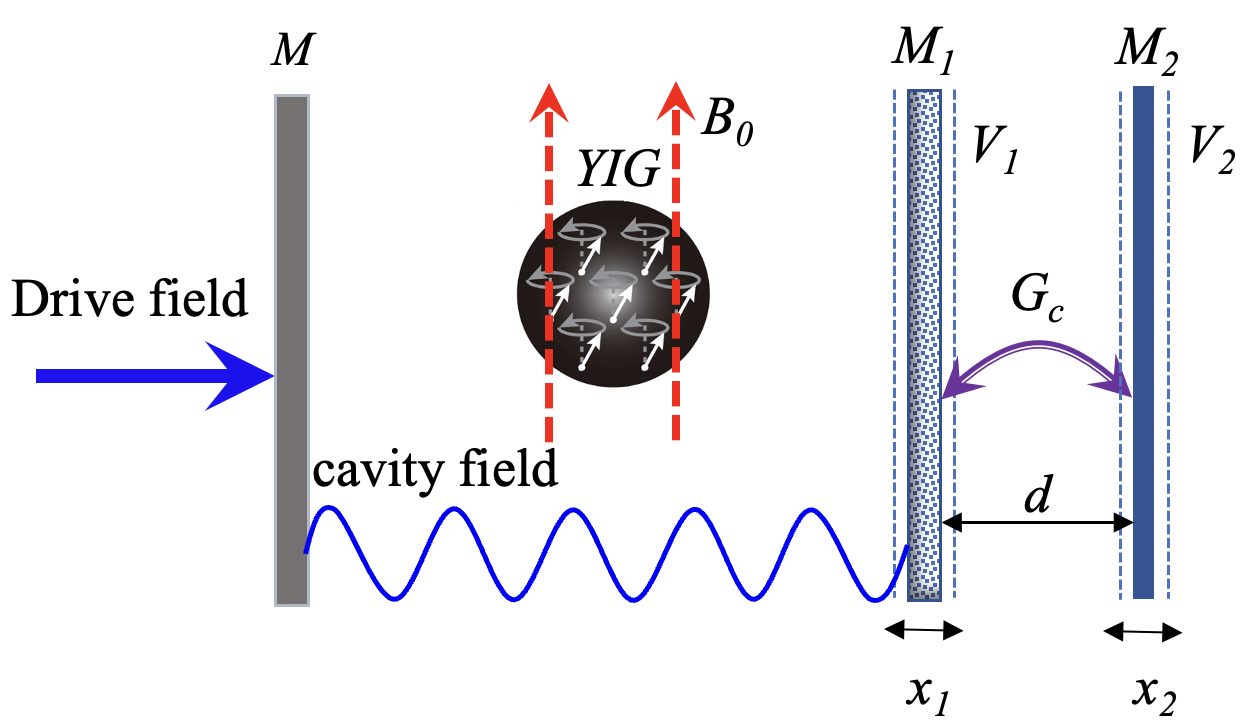}
\caption{The conceptual foundation of the Coulomb-enabled optomechanical cavity magnonic system under study involves a single optomechanical cavity featuring a fixed mirror on the left and a perfectly reflecting mechanical resonator M$_1$ on the right resonating around its equilibrium position, with a Yttrium Iron Garnet (YIG) sphere enclosed between them. The microwave cavity photons, denoted as $c$, couple to the magnon $m$ of the YIG sphere through a magnetic-dipole interaction characterized by coupling strength $g_m$, and to the mechanical resonator M$_1$ via optomechanical coupling with strength $G_0$. The YIG sphere is subjected to an external magnetic drive field $B_0$ with strength $\Omega_B$ to generate an appropriate number of magnons within the YIG sphere.
A third mechanical resonator, M$_2$, on the right interacts with M$_1$ through electrostatic interactions, characterized by coupling strength $G_c$, as both resonators are capacitively charged with opposite polarities via a DC bias voltage. The physical separation between the two resonators is denoted as $d$, while $x_1$ and $x_2$ represent their small displacements from equilibrium positions.
}
\label{fig:model}
\end{figure}

We consider a hybrid cavity optomechanical-magnonic (COMM) system composed of a microwave cavity mode denoted as $c$ magnetically coupled to a magnon mode $m$ of a YIG sphere via magnetic-dipole interaction with coupling strength $g_m$. Additionally, the cavity mode is optomechanically coupled to the motional mode of a charged mechanical oscillator, characterized by its dimensionless position $x_1$ and momentum operator $p_1$, through a radiation pressure interaction with coupling strength $G_0$, as depicted in Fig. \ref{fig:model}. 
The resonators $c$, $m$, and $M_1$ possess resonance frequencies $\omega_c$, $\omega_m$, and $\omega_1$, respectively. Additionally, the mechanical resonator $M_1$ is electrostatically coupled to another mechanical oscillator $x_2$ of the charged movable cantilever $M_2$ via a Coulomb interaction with strength $G_c$, while the parameter $d$ denotes the separation between the two charged mechanical resonators. The motional mode $M_1$ is biased with a voltage denoted by $V_1$, whereas the second resonator $M_2$ is charged by a voltage expressed as $-V_2$.
The YIG sphere is subjected to a strong magnetic field characterized by a strength denoted as $\Omega_B$ (see Appendix A) and a frequency $\omega_B$. This results in magnetic nonlinearity of the magnon, referred to as the magnon self-Kerr effect, which plays a critical role in shifting the magnon resonance frequency. The total Hamiltonian of the hybrid COMM system, as illustrated in Fig. \ref{fig:model}, can be expressed as follows:
\begin{align}
\label{hamiltonian}
H_{T} / \hbar= &  \omega_{c} c^{\dagger} c +\omega_{m} m^{\dagger} m + \frac{1}{2}\sum_{j=1}^{2} \omega_j (x_{j}^2 + p_{j}^2) - G_0 c^{\dagger} c x_1 \nonumber\\[4pt]& + G_c x_1 x_2   + K_s (m^{\dagger} m)^2  + g_{m}\left(c m^{\dagger}+ c^{\dagger} m\right) \nonumber\\[4pt]& + i \Omega_B \left(m^{\dagger} e^{- i \omega_B t} - m e^{ i \omega_B t} \right).
\end{align}
The first two terms of Eq. \eqref{hamiltonian} correspond to the cavity and magnon fields, with $\omega_c$ and $\omega_m$ denoting the resonant frequencies of the cavity and magnon, respectively. The third term describes the mechanical oscillators M$_1$ and M$_2$, also known as harmonic oscillators, where $x_j$ and $p_j$ ($j=1,2$) represent their dimensionless position and momentum operators, respectively. 
The next two terms $G_0 c^{\dagger} c x_1$ and $G_c x_1 x_2$ represent the microwave optomechanical coupling between cavity photons and M$_1$, and the Coulomb coupling between the two mechanical resonators M$_1$ and M$_2$, with $G_0$ and $G_c$ denoting their respective coupling strengths. The Hamiltonian for the Coulomb-coupled M$_1$ and M$_2$ arises from the Coulomb interaction potential, expanded to second order as 
\begin{align}
   G_c = -\frac{C_1 V_1 C_2 V_2}{4\pi \varepsilon_0 d}\left[1 - \frac{x_1 - x_2}{d} + \left(\frac{x_1 - x_2}{d}\right)^2\right],
\end{align}
where $C_i$ and $V_i$ ($i=1,2$) represent the gate capacitance and bias voltage of the gates for M$_1$ and M$_2$, respectively, and $\varepsilon_0$ is the permittivity of free space or the vacuum dielectric constant.
This expression is valid under the condition that the oscillations of the mechanical resonators about their mean positions are significantly smaller ($x_1, x_2 \ll d$) than the separation distance between them. After simplification and renormalization of the resonance frequency for each mechanical resonator, the expression is reformulated as $G_c x_1 x_2$, where $G_c = \frac{C_1 V_1 C_2 V_2}{2 \pi \varepsilon_0 d^3}$ represents the Coulomb coupling strength. The sixth term of the Hamiltonian corresponds to the magnon's intrinsic nonlinear effect, i.e., the self-Kerr nonlinearity, characterized by $K_s$, the magnon's self-Kerr coefficient.
The subsequent beam-splitter-like expression couples the microwave cavity photons with the magnon of the YIG sphere, characterized by the magnetic dipole interaction strength $g_m$. The final term describes the interaction of the magnetic bias field $B_0$ with the magnon of the YIG sphere, with $\Omega_B$  amplitude strength.
By applying the unitary transformation $ U(t) = \exp[-i\omega_{B}(c^{\dagger} c + m^{\dagger} m)t] $ (analogous to the Schrodinger's Equation) in the frame rotating at the magnetic bias frequency $\omega_B$, the system Hamiltonian is transformed, as described by the following expression whose detailed derivation can be found in the Appendix B below.
\begin{align}
\label{hamiltonian1}
H_{T} / \hbar= &  \Delta_{0} c^{\dagger} c +\Delta_{m} m^{\dagger} m + \frac{1}{2}\sum_{j=1}^{2} \omega_j (x_{j}^2 + p_{j}^2) - G_0 c^{\dagger} c x_1 \nonumber\\[4pt]& + G_c x_1 x_2 + K_s (m^{\dagger} m)^2  + g_{m}\left(c m^{\dagger}+ c^{\dagger} m\right) \nonumber\\[4pt]& + i \Omega_B \left(m^{\dagger}-m\right),
\end{align}
where $\Delta_0 = \omega_c - \omega_B$ denotes the cavity-magnetic field detuning, and $\Delta_m = \omega_m - \omega_B$ represents the magnon-magnetic field detuning.
\subsection{Quantum Langevin Equations}

To adhere to the standard approach for open dynamical systems, we introduce noise inputs and dissipation into the model. The system's evolution is subsequently described by the quantum Langevin equations (QLE), which dictate its dynamics in the following form \cite{gardiner2004quantum}:
\begin{equation}
\label{langevin}
 \frac{\mathrm{d} Y}{\mathrm{d}t}=-\frac{i}{\hbar}\left[Y, H_{T}\right]-\Gamma Y+N,
\end{equation}
where $Y$ denotes a general operator variable whose dynamics are being examined, $ \Gamma $ represents the dissipation rate, and $ N $ is the noise operator. By applying the QLE from Eq.  \eqref{langevin} to the frame-rotated Hamiltonian in Eq. \eqref{hamiltonian1}, we derive a set of dynamical equations that include phenomenological relaxation and corresponding noise terms, expressed as:
\begin{subequations}\label{subequations}
\begin{align}
\frac{\mathrm{d}c}{\mathrm{d}t} =& -i \Delta_{0} c + i G_0 c x_1 - i g_m m  - \kappa c + \sqrt{2\kappa} c_{\rm{in}},\label{subeq1}\\[4pt]
\frac{\mathrm{d}m}{\mathrm{d}t} =& -i \Delta_{m} m - i g_m c  - 2i K_s m^{\dagger} m m + \Omega_B  - \gamma_m m \nonumber\\ &+ \sqrt{2\gamma_m} m_{\rm{in}},\label{subeq2}\\[4pt]
\frac{\mathrm{d}x_1}{\mathrm{d}t}  =& \omega_1 p_1,\label{subeq3}\\[4pt]
\frac{\mathrm{d}p_1}{\mathrm{d}t} =& -\hbar \omega_1 x_1 + \hbar G_0 c^{\dagger} c - \hbar G_c x_2 - \gamma_{1} p_1 + \zeta_1,\label{subeq4}\\[4pt]
\frac{\mathrm{d}x_2}{\mathrm{d}t} =& \omega_2 p_2,\label{subeq5}\\[4pt]
\frac{\mathrm{d}p_2}{\mathrm{d}t} =& -\hbar \omega_2 x_2 - \hbar G_c x_1 - \gamma_{2} p_2 +  \zeta_2.\label{subeq6}
\end{align}
\end{subequations}
Here, $\kappa$, $\gamma_m$, and $\gamma_{1,2}$ represent the damping rates of the cavity, magnon, and motional modes, respectively, while $c_{\text{in}}$, $m_{\text{in}}$, and $\zeta_{1,2}$ denote the input noise operators for the cavity and magnon modes, each with zero expectation value, i.e., $\langle c_{\rm{in}}(t) \rangle = \langle m_{\rm{in}}(t) \rangle = 0$. The thermal noise or quantum Brownian stochastic forces acting on the mechanical resonators are also characterized by zero mean values, $\langle \zeta_1(t) \rangle = \langle \zeta_2(t) \rangle = 0$. The equations \eqref{subeq1}-\eqref{subeq6} contain the complete dynamical information about the time evolution of the system's operators. In general, they are coupled nonlinear ordinary differential equations, and analytical solutions are not readily obtainable. 
These stochastic forces in the above equations are inherently non-Markovian. However, when the mechanical quality factor satisfies $Q \gg 1$, they can be approximated as a Markovian process \cite{benguria1981quantum}. Under this assumption, the forces become $\delta$-correlated, such that we have 
\begin{align}\label{stochastic forces}
  \frac{1}{2} \langle \zeta_{1,2} (t) \zeta_{1,2} (t') + \zeta_{1,2} (t') \zeta_{1,2}(t) \rangle \simeq \gamma_{1,2} (2\bar{n}_{1,2} + 1) \delta(t - t').  
\end{align}
The input noise terms of the photon and magnon modes exhibit non-zero correlation functions, respectively, as follows:
    \begin{align}\label{input noise terms}
        &\langle c_{\text{in}}^\dagger(t') c_{\text{in}}(t) \rangle = \bar{n}_c \delta(t - t') &\\[4pt]
        &\langle c_{\text{in}}(t') c_{\text{in}}^\dagger(t) \rangle = (\bar{n}_c + 1) \delta(t - t') &\\[4pt]
        &\langle m_{\text{in}}(t') m_{\text{in}}^\dagger(t) \rangle = (\bar{n}_m + 1) \delta(t - t')&\label{input noise terms1}
    \end{align}
Here, the symbol $\sigma = c, m, 1, 2$ corresponds to the equilibrium mean thermal photon, magnon, or phonon number, whose expression can be given as:
\begin{align} \label{thermal occupation}
    \bar{n}_\sigma = \left[\exp\left(\frac{\hbar\omega_\sigma}{k_B T}\right) - 1\right]^{-1},
\end{align}
where $k_B$ is the Boltzmann constant and $T$ represents the bath or environmental temperature. We assume that the magnons of the YIG sphere are driven by a strong magnetic bias field leading to a large amplitude $|\langle m \rangle| \gg 1$ at the steady-state. Due to the presence of photon-magnon beam-splitter-like interaction, the cavity photons also have large amplitude excitations $|\langle c \rangle| \gg 1$.
Under these conditions, it is reasonable to expand (linearize) the magnon and photon operators as $h = \langle h \rangle + \delta h$, concentrating on steady-state behavior while neglecting higher-order fluctuations in the operators.

The steady-state solutions for all operator variables are obtained by setting the derivative terms in Equations \eqref{subequations}(a) to \eqref{subequations}(f) equal to zero, as shown below:
\begin{align}\label{expectation values}
c_{s} &= \frac{-i g_m m_{s}}{\kappa + i \Delta_c}, &       m_{s} &= \frac{\Omega_B - i g_m c_{s}} {\gamma_m + i \Delta^{\prime}_{m}},\nonumber\\[6pt]
x_{1s} &= \frac{G_0 |c_{s}|^2}{\omega_1 + \frac{\hbar G_c^2}{\omega_2}}  &   x_{2s} &= \frac{\hbar G_c G_0 |c_{s}|^2}{\omega_2 \left(\omega_1 + \frac{\hbar G_c^2}{\omega_2}\right)}.
\end{align}
Here, $\Delta_c := \Delta_0 - G_{\rm eff} x_{1s}$ represents the cavity detuning, which is modified by radiation pressure between the cavity field and the mechanical resonator $M_1$. The effective magnon detuning is given by $\Delta^{\prime}_{m} := \Delta_m + \Delta K$, where $\Delta K$ accounts for the frequency shift of the magnon induced by the Kerr effect of the Kittel mode. The parameter $\Delta K := 2 K_s M$ quantifies the strength of two-magnon interactions, which can facilitate magnon squeezing \cite{PhysRevB.105.245310}. The quantity $M := |m_{s}|^2$ represents the magnon population, which quantifies the \textit{two-magnon scattering effect}, where magnons interact with one another, thereby altering their oscillation frequency. The Kerr nonlinearity $K_s$ is central in shifting or tuning the system's internal dynamics.

To measure the quadrature variance of the photon and magnon amplitudes and phases, we define quadrature operators for both the cavity and magnon modes, where the phase reference of the cavity field is chosen such that $c_s$ is a positive, real-valued quantity. By further defining the quadrature fluctuations, analogous to dimensionless position and momentum, for the cavity and magnon modes as $\delta X_{c} := (\delta c + \delta c^{\dagger}) / \sqrt{2}$ and $\delta Y_{c} := (\delta c - \delta c^{\dagger}) / i \sqrt{2}$ for the cavity mode, and $\delta X_{m} := (\delta m + \delta m^{\dagger}) / \sqrt{2}$ and $\delta Y_{m} := (\delta m - \delta m^{\dagger}) / i \sqrt{2}$ for the magnon mode, along with the corresponding Hermitian input quadratures for the noise operators $\delta X_{c}^{\text{in}} := (c_{\text{in}} + c^{\dagger}_{\text{in}}) / \sqrt{2}$ and $\delta Y_{c}^{\text{in}} := ( c_{\text{in}} - c^{\dagger}_{\text{in}} ) / i\sqrt{2}$ for the cavity, and $\delta X_{m}^{\text{in}} = (m_{\text{in}} + m^{\dagger}_{\text{in}}) / \sqrt{2}$ and $\delta Y_{m}^{\text{in}} := (m_{\text{in}} - m^{\dagger}_{\text{in}}) / i\sqrt{2}$ for the magnon mode, the linearized QLEs can be expressed in a more compact form as follows.
\begin{subequations}
\label{eq:subeq}
\begin{align}
    \delta \dot{x}_1 &= \omega_1 \delta p_1,\label{eq:subeq1}\\[4pt]
    \delta \dot{x}_2 &= \omega_2 \delta p_2,\label{eq:subeq2}\\[4pt]
    \delta \dot{p}_1 &= -\omega_1 \delta x_1 + G_{\rm eff} \delta X_c - G_c \delta x_2 - \gamma_1 \delta p_1 + \zeta_1,\label{eq:subeq3}\\[4pt]
    \delta \dot{p}_2 &= -\omega_2 \delta x_2 - G_c \delta x_1 - \gamma_2 \delta p_2 + \zeta_2,\label{eq:subeq4}\\[4pt]
    \delta \dot{X}_c &= -\kappa \delta X_c + \Delta_c \delta Y_c + g_m \delta Y_m + \sqrt{2\kappa} \delta X_c^{\text{in}},\label{eq:subeq5}\\[4pt]
    \delta \dot{Y}_c =& -\kappa \delta Y_c - \Delta_c \delta X_c + G_{\rm eff} \delta X_1 - g_m \delta X_m \nonumber\\&  + \sqrt{2\kappa} \delta Y_c^{\text{in}},\label{eq:subeq6}\\[4pt]
    \delta \dot{X}_m &= -\gamma_m \delta X_m + \Delta^{\prime}_m \delta Y_m + g_m \delta Y_c + \sqrt{2\gamma_m} \delta X_m^{\text{in}},\label{eq:subeq7}\\[4pt]
    \delta \dot{Y}_m &= -\gamma_m \delta Y_m - \Delta^{\prime}_m \delta X_m  - g_m \delta X_c + \sqrt{2\gamma_m} \delta Y_m^{\text{in}}.\label{eq:subeq8}
\end{align}
\end{subequations}
Here, the term $G_{\rm eff} := \sqrt{2} G_0 c_s$ represents the cavity-enhanced effective optomechanical coupling between the cavity field and the mechanical resonator M$_1$. 

The linearized equations of motion can be expressed in a compact matrix form as follows:
\begin{equation}
    \dot{u}(t) = A u(t) + n(t),
\end{equation}
where the expression 
\begin{align}
 u(t) := (\delta x_1 \ \  \delta p_1 \ \ \delta x_2 \ \ \delta p_2 \ \  \delta X_c \ \ \delta Y_c \ \ \delta X_m \ \ \delta Y_m)^{\mathbf{T}}   
\end{align}
represents a column vector operator of quadrature fluctuations and $\mathbf{T}$ is the transpose operator of the column vector, whereas  
\begin{equation}
\begin{aligned}
n(t) := \Big[ &0 \ \ \zeta_1 \ \ 0 \ \ \zeta_2 \ \sqrt{2\kappa} \ \ \delta X_c^{\text{in}} \ \ 
 \sqrt{2\kappa} \ \ \delta Y_c^{\text{in}}, \\[4pt]& \ \ \ \ \ \ \ \ \ \ \ \ \ \sqrt{2\gamma_m} \ \ \delta X_m^{\text{in}} \ \
 \sqrt{2\gamma_m} \ \ \delta Y_m^{\text{in}} \Big]^{\mathbf{T}}
\end{aligned}
\end{equation}  
corresponds to the transpose of a column vector of the quadrature noise sources.  
Additionally, the $8 \times 8$ drift matrix $A$, which encapsulates the complete characteristics of the quantum steady-state for the fluctuation terms, can be written as:
\begin{equation}\label{drift:matrix}
    A := \begin{pmatrix}
0 & \omega_1 & 0 & 0 & 0 & 0 & 0 & 0 \\
-\omega_1 & -\gamma_1 & -G_c & 0 & G_{\rm eff} & 0 & 0 & 0 \\
0 & 0 & 0 & \omega_{2} & 0 & 0 & 0 & 0 \\
-G_c & 0 & -\omega_2 & -\gamma_2 & 0  & 0 & 0 & 0 \\
0 & 0 & 0 & 0 & -\kappa & \Delta_c & 0 & g_m \\
G_{\rm eff} & 0 & 0 & 0 & -\Delta_c & -\kappa & -g_m & 0 \\
0 & 0 & 0 & 0 & 0  & g_m & -\gamma_m & \Delta^{\prime}_m \\
0 & 0 & 0 & 0 & -g_m & 0 & -\Delta^{\prime}_m & - \gamma_m \\
\end{pmatrix}.
\end{equation}
The matrix $A$ encodes the structural information of the optomechanical magnonic system under investigation and serves as a foundation for further analysis, including the characterization of quantum entanglement, as demonstrated next.

\subsection{Quantum Entanglement Quantification}

Our primary focus is to examine the stationary entanglement between the mechanical modes of the charged mechanical oscillators in this hybrid COMM setup. For the system to be stable and exhibit a unique steady-state solution, the real parts of all eigenvalues of the drift matrix $A$ must be negative. While the stability conditions can, in principle, be derived analytically using the Routh-Hurwitz criterion [64], doing so for our hybrid setup is complex and cumbersome. Therefore, we will employ numerical methods to ensure that the chosen parameters lie within the stable regime.
Since the quantum noises are zero-mean Gaussian and the dynamics are linearized, the stable steady state of the quantum fluctuations in this system is a zero-mean multipartite Gaussian state. This state can be fully characterized by an $8 \times 8$ stationary correlation matrix $V$, with elements defined as:
\begin{align}\label{stationary correlation matrix}
    V_{ij} := \frac{1}{2} \langle u_i(\infty)u_j(\infty) + u_j(\infty)u_i(\infty) \rangle ~ \text{for } i, j = 1, \ldots, 8.
\end{align}
Given that all system parameters meet the stability conditions, the stationary correlation matrix $V$ satisfies the following Lyapunov equation:
\begin{equation}
    A V + V A^{\mathbf{T}} + D= 0.
\end{equation}
where $V$ is the covariance matrix that needs to be computed and $D$ denotes a diagonal matrix known as the diffusion matrix, which characterizes the noise correlations of the motional modes, photon, and magnon modes, as given in Eqs. \eqref{input noise terms}-\eqref{input noise terms1}, and is defined as:

\begin{align}\label{Lyapunov equation}
    D :=& \text{diag}[0, \gamma_1(2\bar{n}_1 + 1), 0, \gamma_2(2\bar{n}_2 + 1), \kappa(2\bar{n}_c + 1),\nonumber\\& \kappa(2\bar{n}_c + 1), \gamma_m (2\bar{n}_m + 1), \gamma_m (2\bar{n}_m + 1)].
\end{align}

In the following, we will utilize the logarithmic negativity to quantitatively assess the level of steady-state entanglement between the two mechanical modes, M$_1$ and M$_2$. To calculate the entanglement of this bipartite subsystem, we need to extract a submatrix from the stationary correlation matrix by excluding the irrelevant degrees of freedom. Initially, we focus on computing the entanglement of the bipartite subsystem composed of the two Coulomb-coupled mechanical modes; this requires obtaining a $4 \times 4$ submatrix $\Psi$ from the $8 \times 8$ stationary correlation matrix $\mathbf{V}$ by disregarding the remaining unconcerned degrees of freedom.
In this manner, the reduced $4 \times 4$ correlation matrix $\Psi$ for a two-mode Gaussian state in the continuous variable system can be succinctly represented as a $2 \times 2$ block matrix:
\begin{equation}
\Psi = \begin{pmatrix} 
B_1 & E \\[4pt]
E^{\mathbf{T}} & B_2 
\end{pmatrix},
\end{equation}
where $B_1$ and $B_2$ are the sub-block matrices representing the local properties of M$_1$ and M$_2$, respectively, while $E$ describes the non-local correlation between these two mechanical modes. A similar approach is employed to extract the reduced correlation matrix for the entanglement between cavity photons and the phonons of the mechanical resonator M$_1$. The expression for the quantum entanglement, described in terms of logarithmic negativity, can be written as:
\begin{equation}\label{entanglement}
E_N = \max \left[ 0, -\ln 2\eta^- \right],
\end{equation}
where we have
$$\eta^- := 2^{-1/2} \left[ \Sigma - \left( \Sigma^2 - 4 \det \Psi \right)^{1/2} \right]^{1/2},$$ 
while
$$\Sigma := \det B_1 + \det B_2 - 2 \det E$$ 
is the minimum symplectic eigenvalue of the partial transpose of the reduced $4 \times 4$ submatrix $\Psi$. The two mechanical modes are considered entangled if and only if the smallest symplectic eigenvalue satisfies the condition $\eta^{-} < 0.5$, which corresponds to $E_N > 0$. Consequently, we numerically illustrate the results of the analytical expression for the entanglement given in Eq. \eqref{entanglement} through detailed examples in the next section, utilizing experimentally feasible parameters.

\section{Results and Discussion}
\label{results}
In this section, we present the numerical results and discuss the bipartite continuous variable entanglement between the two mechanical oscillators in our proposed hybrid optomechanical system. We will analyze the entanglement characteristics under various parameter settings and explore the impact of different system configurations on the entanglement dynamics. To ensure that the numerical results are reasonable and closely aligned with current experimental conditions \cite{zhang2014strongly,tabuchi2014hybridizing,zhang2016cavity}, we select the following relevant optomechanical parameters:
$\omega_1/2\pi = \omega_2/2\pi = 10 ~\text{MHz}$, $\omega_m /2\pi =  10 ~\text{GHz}$, $\gamma_m/2\pi = 0.1 ~\text{MHz}$, $\kappa /2\pi = 3 ~\text{MHz}$, 
$\gamma_1 /2\pi = \gamma_2/2\pi = 200 ~\text{Hz}$, $G_{\text{eff}} = 0.55 \kappa$,
$\Delta_m = \omega_1$,
$\Delta K = 0.65 \omega_1$. Note that 
$\hbar = 1.0546 \times 10^{-34} ~\text{J}\cdot\text{s}$ and the bath temperature is set at
$T = 10 ~ \text{mK}$.
\subsection{Quantifying Entanglement $E_N$ Between M$_1$ and M$_2$ by Tuning Coulomb Coupling}
\label{results:M1 and M2 vs Gc}
\begin{figure*}
\centering
  \includegraphics[width=0.85\linewidth]{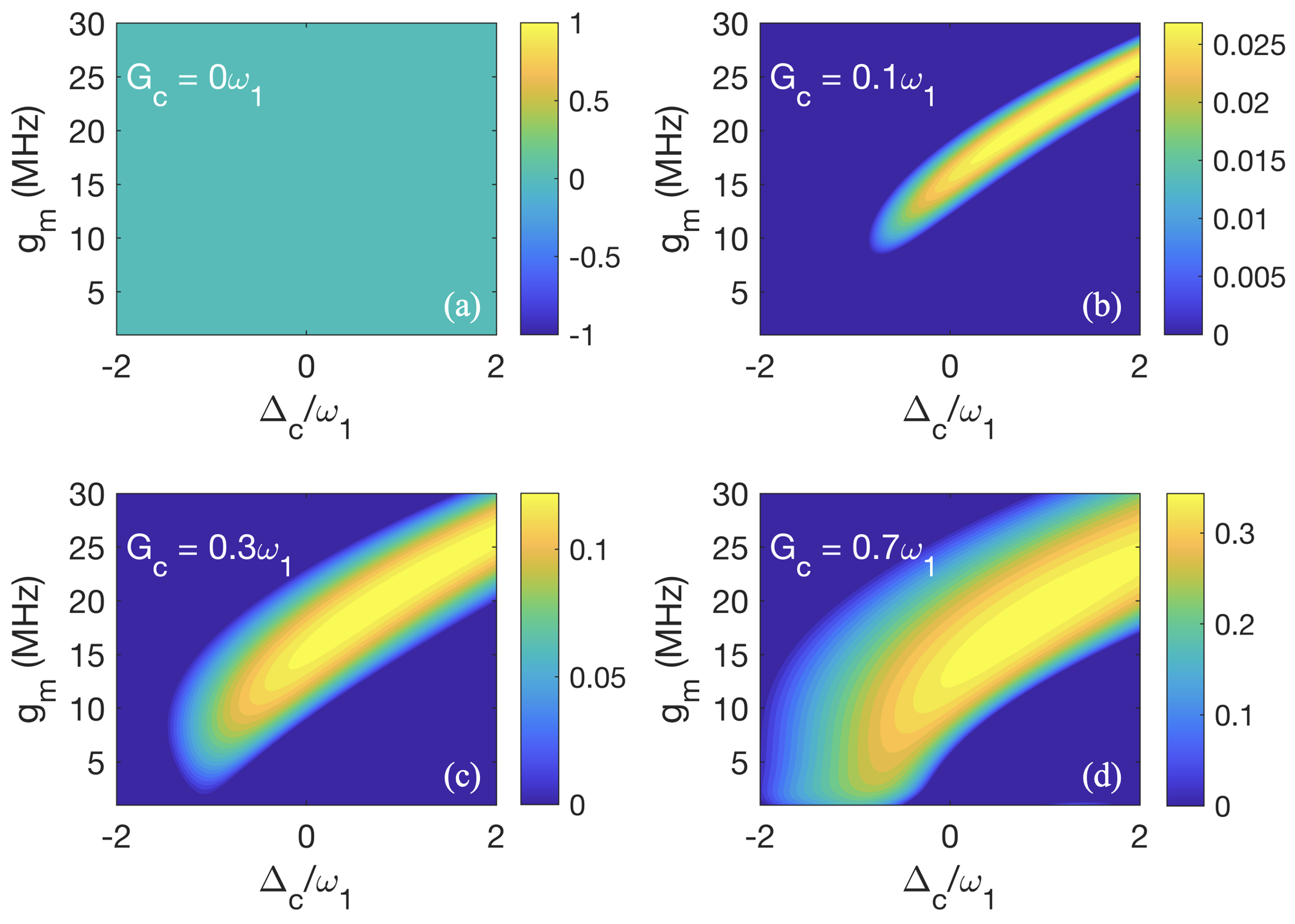}
  \caption{(Color online) Steady-state quantum entanglement $E_N$ between two charged mechanical resonators plotted against scaled cavity detuning $\Delta_c$ and magneto-optical coupling strength under different values of Coulomb coupling (a) $G_c = 0$, (b) $G_c = 0.1 \omega_1$, (c) $G_c = 0.3 \omega_1$, and (d) $G_c = 0.7 \omega_1$. The general system parameters are: $\omega_1/2\pi = \omega_2/2\pi = 10~ \text{MHz}$, $\omega_m /2\pi = 10 ~\text{GHz}$, $\gamma_m/2\pi = 0.1 ~\text{MHz}$, $\kappa /2\pi = 5.5 ~\text{MHz}$, $\gamma_1 /2\pi = \gamma_2/2\pi = 200 ~\text{Hz}$, $G_{\text{eff}} = 0.55 \kappa$, $\Delta_m = \omega_1$, $\Delta K = 0.65 \omega_1$,  $\hbar = 1.0546 \times 10^{-34} ~\text{J}\cdot\text{s}$, and the bath temperature $T = 10~ \text{mK}$.}
  \label{GcControlledEN}
\end{figure*}

As the proposed hybrid COMM system is influenced by the nonlinearity of the magnon, we investigate the entanglement between two spatially separated mechanical resonators via electrostatic interaction under the influence of magnon-optic coupling $g_m$. The electrostatic force, or Coulomb coupling strength, acting between the charged mechanical resonators facilitates direct interaction, paving the way for quantum correlations to develop between their motional states. This coupling mechanism provides an exceptional environment for exploring continuous-variable entanglement, where the strength of the Coulomb interaction plays a pivotal role in examining the entanglement dynamics in our setup.

Figure \ref{GcControlledEN} illustrates the behavior of quantum entanglement between the motional modes of mechanical resonators M$_1$ and M$_2$ as the Coulomb coupling strength $G_c$ is varied. As evident from Eqs. \eqref{drift:matrix} and  \eqref{entanglement}, there is no interaction between the two charged resonators in the absence of the Coulomb coupling $G_c$.
In Fig. \ref{GcControlledEN}(a), no signature of entanglement is observed when $G_c = 0$, regardless of the photon-magnon interaction strength. However, upon tuning $G_c$ to a non-zero value, a contour emerges, indicating the presence of entanglement between the two motional modes, as illustrated in Fig. \ref{GcControlledEN}(b). The degree of entanglement progressively increases with the enhancement of the Coulomb interaction. This trend is intuitive, as a stronger Coulomb interaction between the charged resonators results in greater and more robust entanglement, as shown in Figs. \ref{GcControlledEN}(c) and \ref{GcControlledEN}(d).
The value of entanglement increases from $E_N = 0$ to approximately $E_N \approx 0.3$, accompanied by a broadening of the entanglement contour. This indicates that entanglement is sustained over a wider frequency range. Notably, as $G_c$ is tuned to larger values, we observe that the contour encompasses lower values of the magneto-optical coupling strength $g_m$. This shows that the stronger Coulomb interaction significantly affects the optomechanical coupling $G_{\rm{eff}}$, leading to modifications in the resonance conditions of the cavity and consequently altering the dynamics of the system.

The above results suggest that the magnon and Coulomb interactions influence each other indirectly. Consequently, the tuning of $G_c$ is related to the resonance frequency of mechanical resonator M$_1$, where larger values (i.e., $G_c \geq \omega_1$) can lead to instability in the system. Thus, it is evident that the quantum entanglement between the two mechanical resonators M$_1$ and M$_2$ strongly depends on the Coulomb coupling strength, which contributes to the generation and robustness of quantum entanglement.
\subsection{Influence of Thermal Bath $T$ on $E_N$ Between Mechanical Resonators}
\label{M1 M2 vs bath temperature}
\begin{figure*}[tbhp]
\centering
  \includegraphics[width=0.9\linewidth]{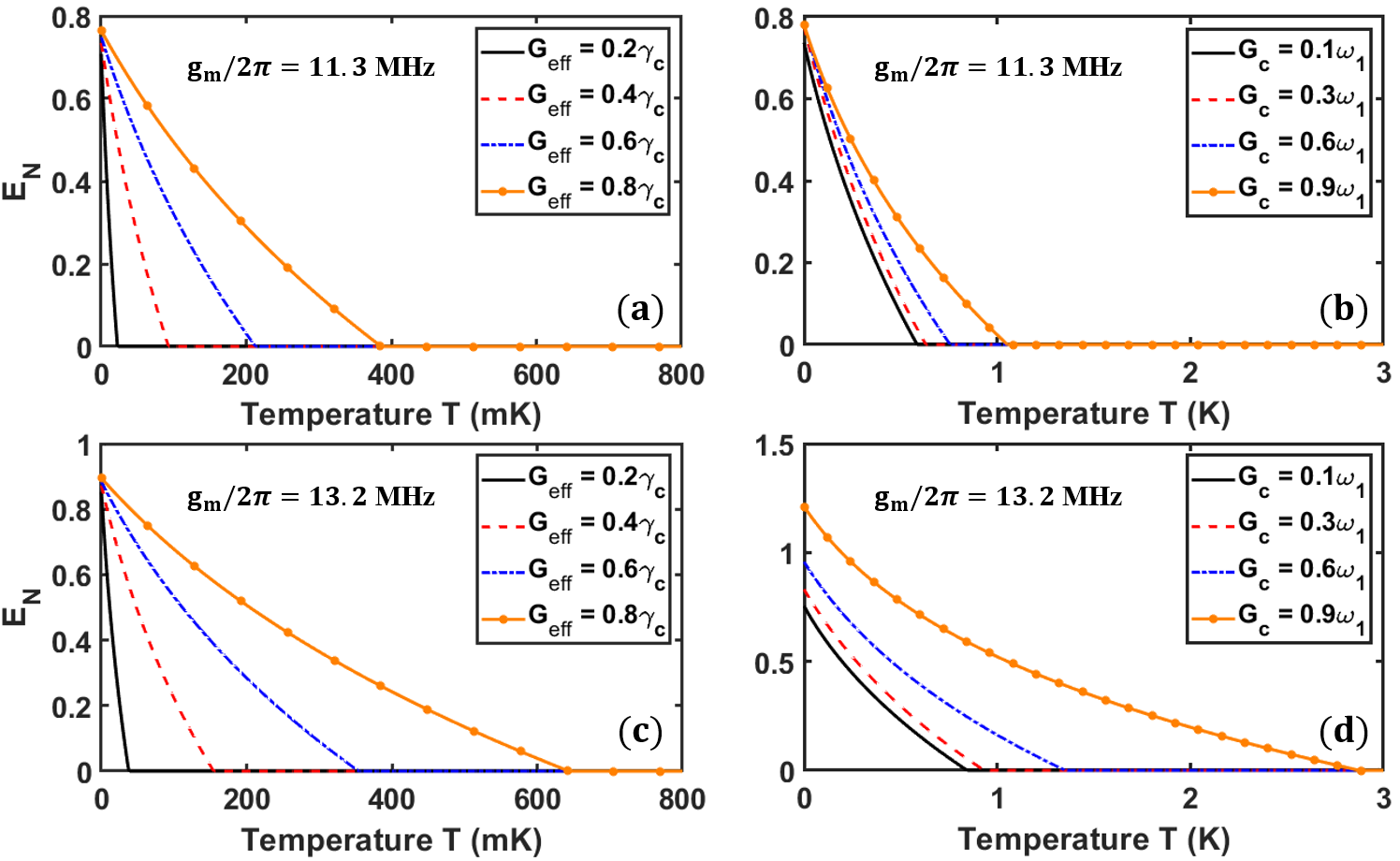}
  \caption{(Color online) Entanglement ($E_N$) between two charged mechanical resonators plotted against bath temperature $T$, (a) and (b) under different values of optomechanical coupling $G_{\rm eff}$ and Coulomb coulomb coupling $G_c$ when $g_m /2\pi = 11.3$ MHz, (c) and (d) under different values of $G_{\rm eff}$ and $G_c$, respectively, when $g_m /2\pi = 13.2$ MHz. The general parameters used are: (a), (c) $G_c = 0.5\omega_1$, and (b), (d) $G_{\rm{eff}} = 0.95\kappa$. Other parameters used here are the same as in Fig. \ref{GcControlledEN}.}
  \label{fig:ENvsM1M2vsT}
\end{figure*}
\begin{figure}
\centering
\includegraphics[width=0.95\linewidth]{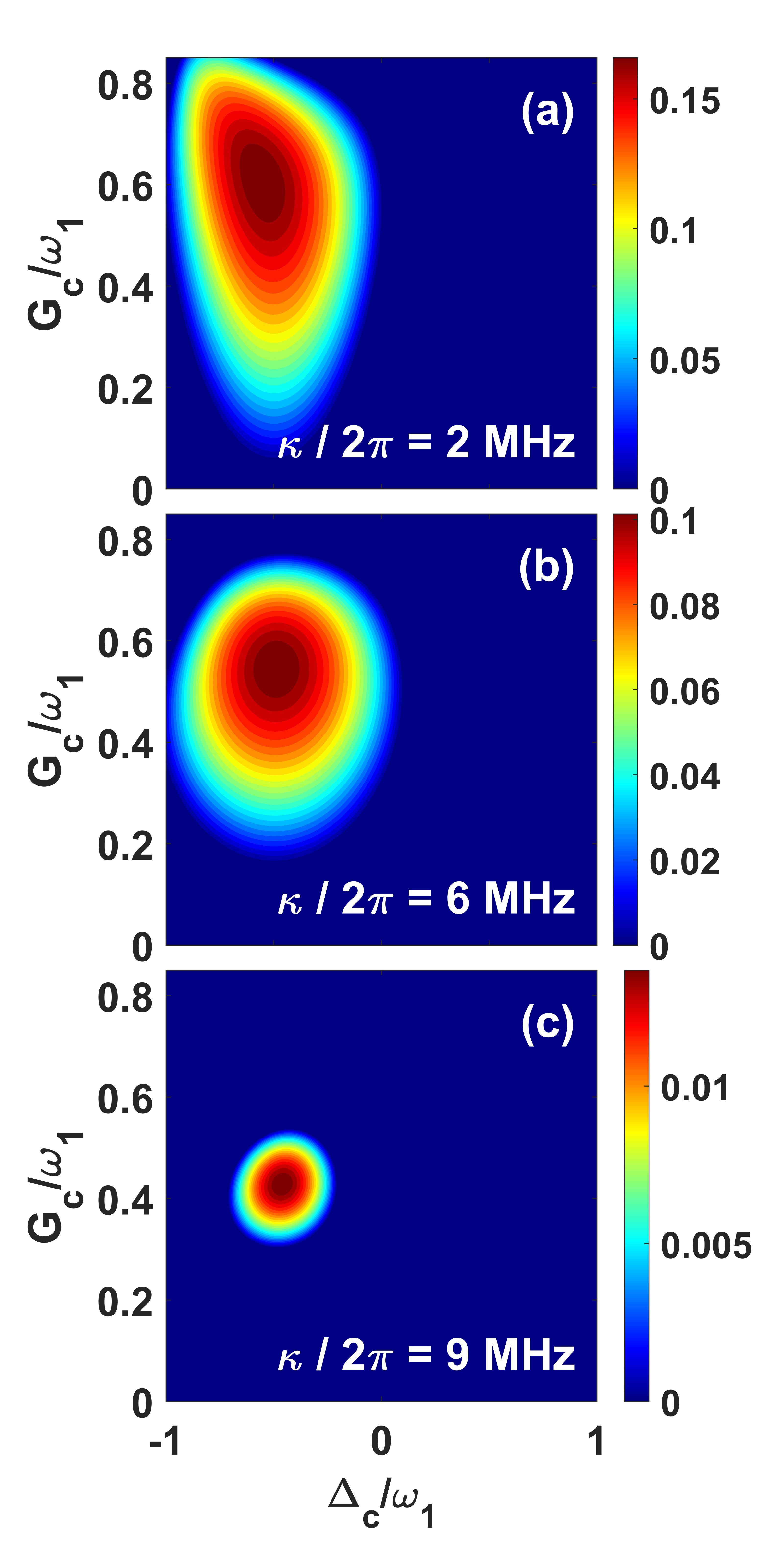}
  \caption{(Color online) Contour plot of Entanglement ($E_N$) between two motional modes plotted against scaled cavity detuning and Coulomb coupling interaction under different values of photon loss rate parameter: (a) $\kappa /2\pi = 2$ MHz, (b) $\kappa /2\pi = 6$ MHz, and (c) $\kappa /2\pi = 9$ MHz. The general parameters used are: $G_{\rm{eff}}/2\pi = 1.5$ MHz, $g_m /2\pi = 10$ MHz, and $ T = 15$ mK. Other parameters used here are the same as in Fig. \ref{GcControlledEN}.}
  \label{fig:decay rate}
\end{figure}
\begin{figure}
\centering
\includegraphics[width=\linewidth]{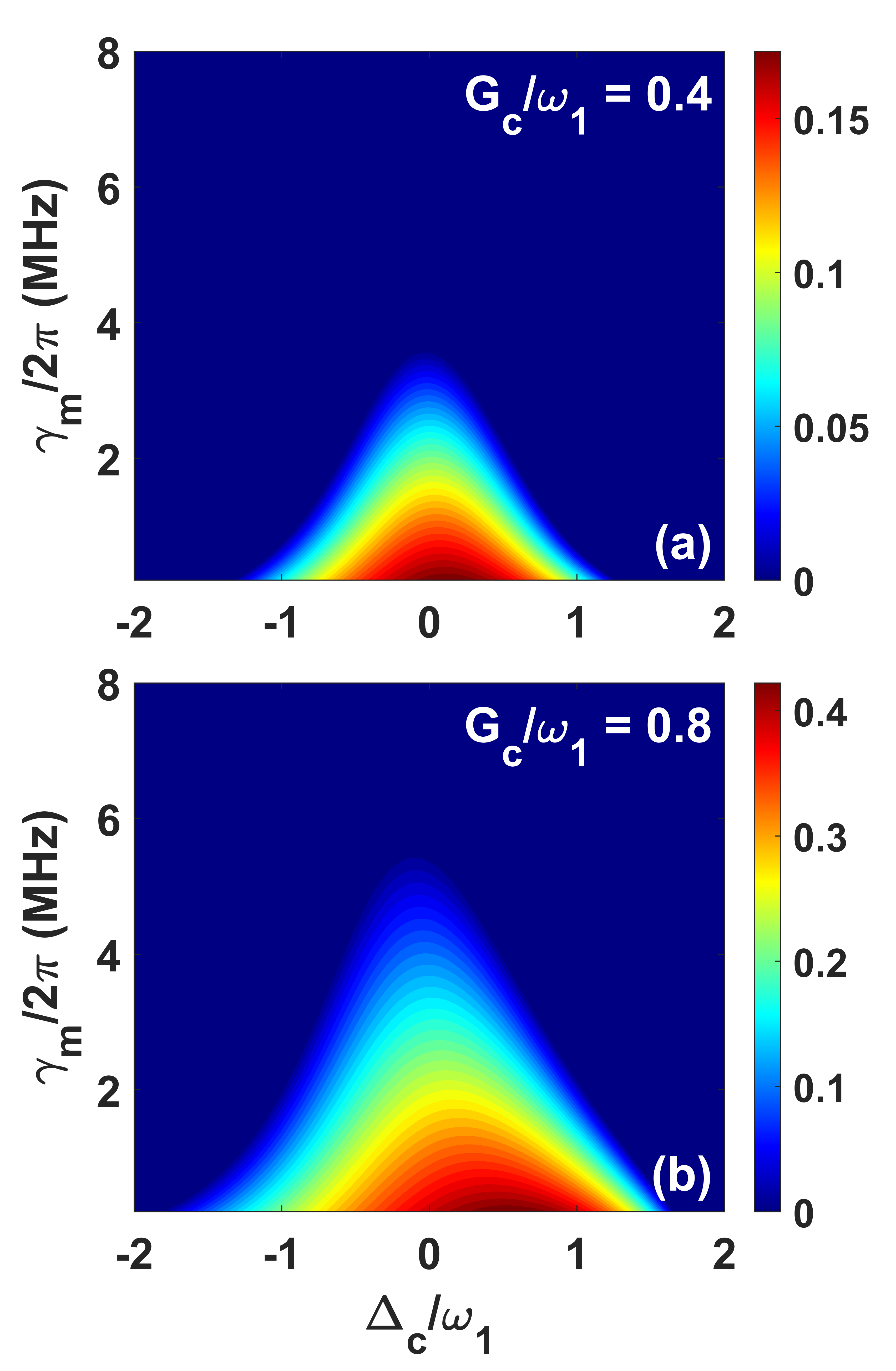}
  \caption{(Color online) Contour plot for the quantum Entanglement between two charged mechanical resonators plotted against the scaled cavity detuning $\Delta_c$ and the magnon loss rate $\gamma_m$. (a) when the Coulomb coupling strength is $G_c = 0.4\omega_1$, and (b) the Coulomb coupling strength is tuned to $G_c = 0.8\omega_1$. The general parameters used are: $g_m /2\pi = 15.5$ MHz. The other parameters used in this figure have the same values as depicted in Fig. \ref{GcControlledEN}.}
  \label{magnon loss}
\end{figure}
It is widely recognized that natural decoherence resulting from strong interactions with the external environment renders the production of non-classical effects highly susceptible to thermal noise. This sensitivity is particularly pronounced in optomechanical systems, where the internal mechanical motions are directly coupled to an external thermal bath. This direct coupling amplifies their vulnerability to fluctuations in environmental temperature.
Therefore, it is crucial to investigate whether the mechanical entanglement that has been achieved can withstand the influences of environmental temperature. In our proposed COMM system, the presence of a magnetic material (YIG) sphere within the cavity setup plays a significant role in enhancing the robustness of the mechanical entanglement against environmental temperature $T$.

Here, we discuss three major parameters: the optomechanical coupling $G_{\rm eff}$, the electrostatic interaction $G_c$, and the magneto-optical interaction $g_m$, and their impact on the robustness of entanglement against the external bath. Initially, we set the value of the magneto-optical interaction to $g_m/2\pi = 11.2$ MHz and examine the mechanical entanglement in relation to external temperature for various values of optomechanical coupling.
At lower values of optomechanical coupling, specifically $G_{\rm eff} = 0.1\kappa$, we observe fragile entanglement that can endure up to approximately 6 milliKelvin (mK) on the temperature scale, as shown in {Fig. \ref{fig:ENvsM1M2vsT}(a)}. When tuning to a higher value of $G_{\rm eff} = 0.8\kappa$, a significantly greater level of entanglement (around 75 mK) is achieved. This indicates that a stronger optomechanical interaction enhances the correlation between the two mechanical resonators, leading to a more robust quantum entanglement of their respective modes. 
Now that both mechanical resonators are coupled via electrostatic interaction, it is clear that the presence of $G_c$ can influence the resonance frequency of M$_1$, thereby modifying the dynamics of the hybrid system. To evaluate its impact on $E_N$, we maintain a constant optomechanical coupling while varying the Coulomb coupling $G_c$ under the same magneto-optical interaction ($g_m/2\pi = 11.2$ MHz).

Interestingly, the presence of $G_c$ can extend the duration of entanglement to higher values of environmental temperature. Specifically, even a lower value of Coulomb coupling (here, $G_c = 0.1\omega_1$) allows entanglement to remain active at approximately 470 mK, as illustrated in Fig. \ref{fig:ENvsM1M2vsT}(b). Larger values of electrostatic coupling facilitate stronger interactions between the two charged mirrors, resulting in more robust and long-lived entanglement.
In Fig. \ref{fig:ENvsM1M2vsT}(b), we observe significant robustness of entanglement at an environmental temperature of approximately 875 mK when the Coulomb coupling is tuned to $G_c = 0.9\omega_1$, while ensuring the stability conditions of the system. Next, we increase the magneto-optical coupling to a higher value of $g_m/2\pi = 14.0$ MHz and examine the behavior of entanglement against the external bath temperature $T$ for the same values of optomechanical and Coulomb coupling strengths as indicated in Figs. \ref{fig:ENvsM1M2vsT}(a) and \ref{fig:ENvsM1M2vsT}(b).

The entanglement between the two mechanical resonators exhibits excellent robustness against environmental temperature in both cases: varying values of $G_{\rm eff}$, as shown in Fig. \ref{fig:ENvsM1M2vsT}(c), and a range of values for $G_c$, presented in Fig. \ref{fig:ENvsM1M2vsT}(d). When the optomechanical coupling is set to $G_{\rm eff} = 0.8\kappa$, we observe a significant enhancement in the robustness of entanglement between the mechanical resonators, as illustrated in Fig. \ref{fig:ENvsM1M2vsT}(c).
The system's robustness against environmental temperature has nearly doubled, now withstanding an external bath of 147 mK, compared to the previous limit of 75 mK. When the Coulomb coupling is set to $G_c = 0.9\omega_1$, the interaction between the motional modes of the charged mechanical resonators becomes significantly stronger, particularly due to the increased photon-magnon coupling. This enhanced interaction results in remarkable resilience of entanglement in the presence of an external bath, as illustrated in Fig. \ref{fig:ENvsM1M2vsT}(d).
The entanglement in this case can withstand a temperature of $ T = 2.8 \, \text{K} $, which is double that observed in Fig. \ref{fig:ENvsM1M2vsT}(b). When cavity photons couple efficiently to the magnon of the YIG sphere, it leads to the hybridization of optical and magnon modes, resulting in the formation of hybrid photon-magnon modes. These hybrid modes modify the energy exchange within our proposed COMM setup, giving rise to new resonance frequencies and altering the behavior of both the cavity photons and mechanical resonators. Consequently, the inclusion of YIG and its excitation within the cavity setup indirectly enhances the overall entanglement in the system by creating additional pathways for quantum correlations among various optical, mechanical, or hybrid modes.
\subsection{Entanglement vs. Cavity Decay and Magnon Loss Rates ($\kappa$ and $\gamma_m$)}
\label{cavity decay rate}
Quantum entanglement between the charged mechanical resonators in our setup is facilitated by radiation pressure forces exerted by photons on M$_1$ within the optical cavity. When the cavity decay rate $\kappa$ is low, the effective optomechanical coupling between the cavity and the mechanical resonator M$_1$ becomes sufficiently strong, resulting in enhanced entanglement between the charged mechanical resonators, as illustrated in Fig. \ref{fig:decay rate}(a).
The hybrid COMM system achieves an optimal level of entanglement (0.2 in this case) between the two mechanical resonators when the cavity decay rate is set to $\kappa /2\pi = 2$ MHz, as indicated by the broad bright contour in Fig. \ref{fig:decay rate}(a). Additionally, larger values of Coulomb interaction contribute to stronger correlations between the mechanical resonators. Conversely, in a dissipative system, as the cavity decay rate increases, the cavity-enhanced optomechanical coupling diminishes, leading to a gradual reduction in the optical field's effectiveness inside the cavity to mediate entanglement, as shown in Fig. \ref{fig:decay rate}(b). This behavior can be attributed to the cavity's inability to retain photons for extended periods, thereby weakening the establishment of strong quantum correlations due to a significantly reduced photon lifetime within the cavity.
Fig. \ref{fig:decay rate}(c) illustrates a significant reduction in entanglement as the cavity decay rate increases to $\kappa /2\pi = 9$ MHz. The density plot indicates that the brighter effective area corresponding to possible entanglement has been considerably diminished compared to the scenarios with lower cavity decay rates. Notably, even stronger Coulomb interaction fails to sustain robust entanglement at such elevated decay rates. A high cavity loss rate exacerbates the decoherence rate of the hybrid COMM setup, as photons trapped within the cavity escape into the environment. This decoherence process degrades quantum information and further diminishes the entanglement between the mechanical resonators.

Since the proposed hybrid cavity optomechanical setup is dominantly influenced by the magnons of the YIG sphere, the decay or loss rate of the magnon promptly plays a crucial role in changing the behavior of entanglement between the two mechanical resonators. Figure \ref{magnon loss} illustrates the impact of the magnon loss rate $\gamma_m$ in quantifying the quantum entanglement. In Fig. \ref{magnon loss}(a), we observe the entanglement behavior between M$_1$ and M$_2$ upon increasing values of magnon loss rate $\gamma_m$. It can be seen that as the $\gamma_m$ value rises, the quantity of entanglement slowly fades out, and vanishes completely at $\gamma_m /2\pi = 3.5$ MHz when the Coulomb coupling strength is tuned to $G_c = 0.4 \omega_1$.

We already know that the entanglement between mechanical resonators is influenced directly by $G_c$ and the optomechanical coupling $G_{\text{eff}}$, which is affected by the magnon loss rate, we tune the Coulomb coupling to $G_c = 0.8 \omega_1$ to see the impact of $\gamma_m$ on $E_N$. Due to the strong Coulomb interaction between M$_1$ and M$_2$, the quantum entanglement survives against a relatively larger value of magnon loss rate as shown in Fig. \ref{magnon loss}(b). Under this value of Coulomb coupling, the quantum entanglement vanishes at $\gamma_m /2\pi = 5.4$ MHz.

Hence, we can clearly find out that the increasing loss rate of magnons inside the cavity leads to a lower influence of magnons on the mechanical resonator M$_1$ and thus results in the loss of correlation between the two motional modes.
\subsection{Sensitivity of Entanglement Against Magnon Detuning}
\label{magnon detuning}
\begin{figure*}
\centering
\includegraphics[width=0.90\linewidth]{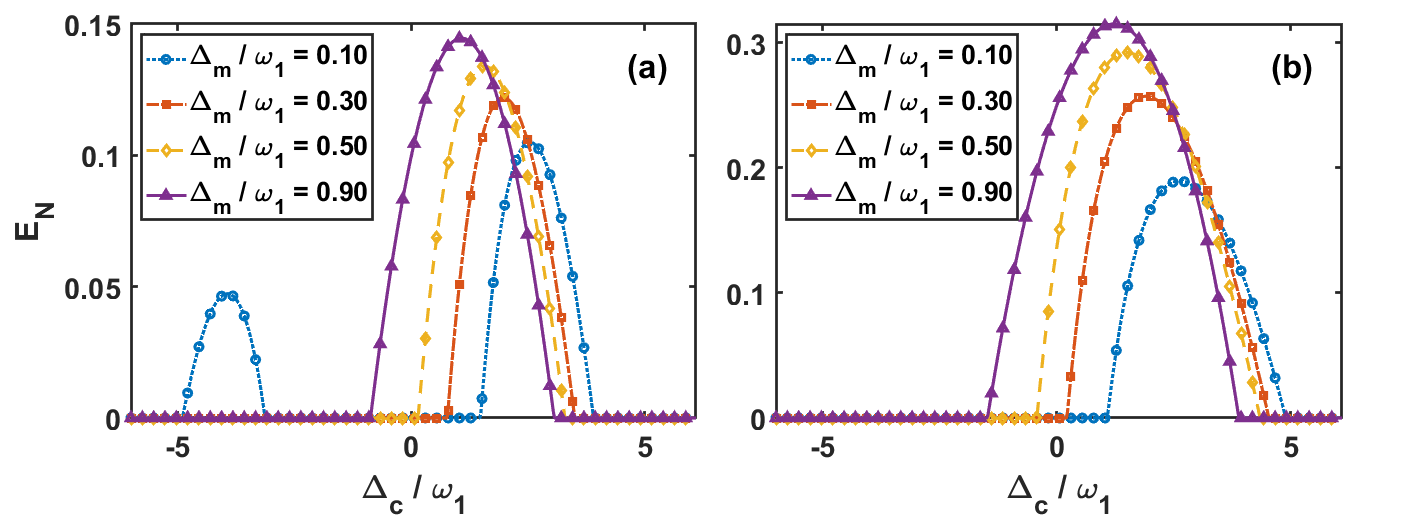}
  \caption{(Color online) Quantity of entanglement between two charged mechanical resonators plotted against the scaled cavity detuning $\Delta_c$ under different values of scaled magnon detuning $\Delta_m$, (a) when the Coulomb coupling strength is $G_c = 0.3\omega_1$, and (b) the Coulomb coupling strength is tuned to $G_c = 0.6\omega_1$. The general parameters used are: $g_m /2\pi = 20$ MHz, $G_{\rm{eff}} = 0.6\kappa$, and $\Delta K = 0.1 \omega_1$. Other parameters used have the same values as depicted in Fig. \ref{GcControlledEN}.}
  \label{fig:magnondetuning}
\end{figure*}

While the magnon-magnetic-drive field detuning $\Delta_m$ influences the effective interaction strength of the cavity field, the selection of its value is crucial and requires careful consideration. Drawing inspiration from cavity optomechanical experiments \cite{brooks2012non}, we utilize a slight red detuning ($\Delta_m < \omega_1$). This choice is informed by the fact that red detuning promotes system stability by facilitating magnomechanical cooling of the phonon mode, thereby enhancing system performance \cite{li2023squeezing}, maximizing coherence, and stabilizing the coupling between the two charged mechanical resonators. Specifically, we examine the entanglement of the mechanical resonators under various parameter values of $\Delta_m$.
Since the entanglement between the two charged resonators depends critically on their electrostatic interaction, we present two examples to illustrate the influence of Coulomb coupling strength on both the degree and behavior of entanglement. Figure \ref{fig:magnondetuning}(a) displays the entanglement measure $E_N$ at a lower Coulomb coupling strength of $G_c = 0.3 \omega_1$. The blue curve indicates a low magnitude of entanglement at two distinct points along the cavity detuning spectrum, $\Delta_c$, when the magnon detuning is absent (i.e., $\Delta_m = 0$).

The introduction of magnon detuning into the setup shifts the magnon mode relative to the cavity field, imparting nonlinearity to the system that is crucial for enhancing quantum correlations. In the absence of $\Delta_m$, the effective coupling of the magnon to the cavity field remains weak, thereby reducing the capacity to sustain entanglement. However, as the detuning increases to a small non-zero value of $\Delta_m = 0.3\omega_1$, the amount of entanglement begins to increase, reaching $E_N = 0.1$, as indicated by the red dashed curve in Fig. \ref{fig:magnondetuning}(a). 
By further tuning the detuning value closer to the resonance frequency of mechanical resonator M$_1$ ($\Delta_m = 0.9\omega_1$), we observe a significant increase in both the magnitude and width of the entanglement. To build upon the role of electrostatic interactions, the Coulomb coupling is set to $G_c = 0.6 \omega_1$ to investigate the entanglement behavior between the two charged motional modes across various magnon detuning values, as previously presented in Fig. \ref{fig:magnondetuning}(a). With this stronger Coulomb coupling, $G_c$, a marked enhancement in the magnitude of entanglement is observed, as depicted in Fig. \ref{fig:magnondetuning}(b).
 The entanglement curve for $\Delta_m = 0$ in Fig. \ref{fig:magnondetuning}(b) closely resembles that for $\Delta_m = 0.9\omega_1$, as shown in Fig. \ref{fig:magnondetuning}(a). As the magnon detuning $\Delta_m$ approaches the frequency of M$_1$, the entanglement curves continue to expand, ultimately surpassing a magnitude of $0.3$ when the magnon detuning value is very close to mechanical resonance, specifically at $\Delta_m = 0.9\omega_1$. 

Another notable observation from Fig. \ref{fig:magnondetuning}(b) is that with the stronger electrostatic interaction $G_c$ between the two charged motional modes, the width of the entanglement $E_N$ curve also broadens significantly, as indicated by the magenta-colored curve at $\Delta_m = 0.9\omega_1$. This curve extends along the cavity detuning range, surpassing the detuning limits of $\Delta_c /\omega_1 = -2$ and $4$.
 For the same magnon detuning value ($\Delta_m = 0.9\omega_1$), lower values of Coulomb coupling lead to a narrower width of the entanglement curve, which remains confined within the cavity detuning range of $\Delta_c /\omega_1 = -2$ and $4$, as illustrated in Fig. \ref{fig:magnondetuning}(a).
 This expansion of the curves suggests that a stronger Coulomb coupling not only enhances the magnitude of entanglement but also sustains it over a wider range of cavity detuning values. As the curve widens, the entanglement becomes less sensitive to precise tuning conditions, effectively stabilizing the entangled state across a broader frequency spectrum. This behavior is particularly advantageous in practical applications, indicating that the system can maintain high levels of entanglement even with minor fluctuations in detuning, which may arise from environmental or operational variations. Thus, increasing $G_c$ not only amplifies entanglement but also enhances its robustness under different operating conditions.
\subsection{Influence of $G_c$ and $g_m$ on Entanglement Between Cavity Photon and Motional Mode  M$_1$}
\label{results:cavity and M1 vs Gc}
\begin{figure*}
\includegraphics[width=0.95\linewidth]{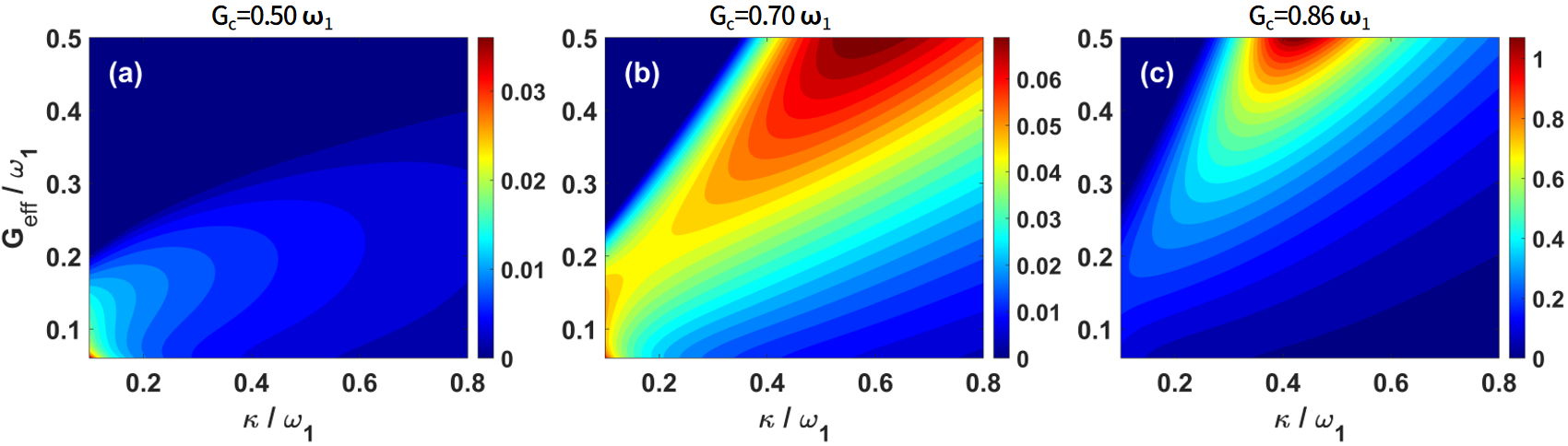}
  \centering
  \caption{Quantum entanglement between cavity photon mode and mechanical resonator M$_1$ motional mode plotted against the scaled cavity decay rate $\kappa$ under different values of Coulomb coupling  (a) $G_c = 0.50 \omega_1$, (b) $G_c = 0.70 \omega_1$, and  (c) $G_c = 0.86 \omega_1$. The general parameters used are $g_m /2\pi= 14.5$ MHz, $\Delta_c = \omega_1$, and $\Delta_m = 2 \omega_1$. Other parameters have the same values as depicted in Fig. \ref{GcControlledEN}.}
  \label{fig:cavEnM1Gc}
\end{figure*}
\begin{figure*}
\includegraphics[width=\linewidth]{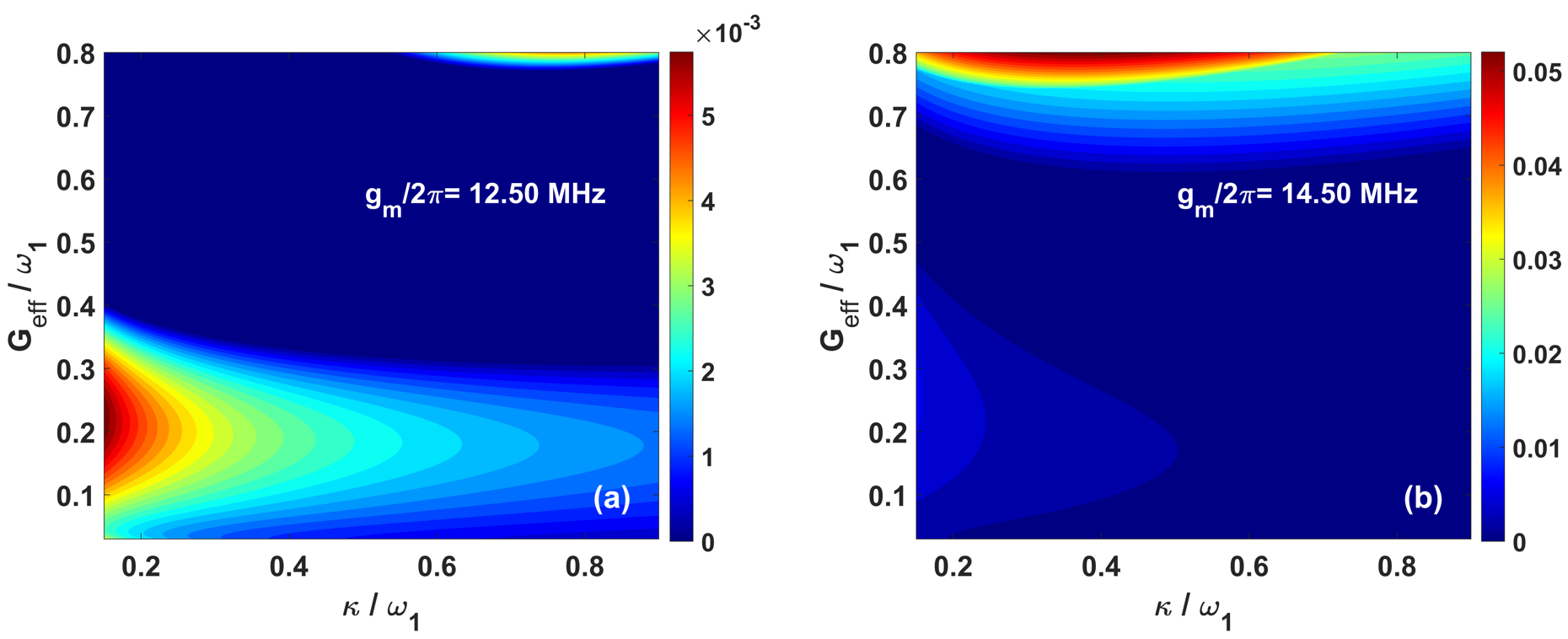}
  \centering
  \caption{Quantum entanglement between cavity photon mode and mechanical resonator M$_1$ motional mode plotted against the scaled cavity decay rate $\kappa$ under different values of magneto-optical coupling $g_m$: (a) $g_m /2\pi = 12.50$ MHz, and (b) $g_m /2\pi = 14.50$ MHz. The value of the Coulomb coupling parameter used here is given as   $G_c = 0.6 \omega_1$. Other parameters have the same values as depicted in Fig. \ref{fig:cavEnM1Gc}.}
  \label{fig:cavEnM1gm}
\end{figure*}
In a cavity optomechanical system, quantum entanglement between the photons of the cavity field and the phonons of the mechanical resonator arises from the radiation pressure interaction, or optomechanical interaction, which couples the optical and mechanical modes. This interaction generates a non-trivial correlation between the photon state in the cavity and the position and momentum of the mechanical resonator. As a result, the state of the cavity field becomes entangled with the motional state of the mechanical resonator.
Typically, the degree and magnitude of quantum entanglement can be engineered and manipulated by controlling various system parameters, such as cavity detuning, mechanical frequency, and coupling strength. In contrast to previous studies \cite{liu2023enhanced,mekonnen2024boosting}, our proposed hybrid COMM system investigates the entanglement between the cavity field and the motional mode of M$_1$ by varying the Coulomb coupling to different values. We posit that $ G_c $ can effectively alter the dynamics of the mechanical resonator, significantly modifying and enhancing the robustness of the entanglement. Furthermore, the presence of the YIG sphere has a substantial impact on the cavity dynamics, playing a crucial role in modulating the correlation between the cavity and motional modes.

First, we examine the entanglement by varying the Coulomb coupling interaction. Lower values of $G_c$ yield minimal levels of entanglement, as illustrated in {Fig. \ref{fig:cavEnM1Gc}}. It is important to note that even in the absence of Coulomb coupling, the phenomenon of entanglement can still be realized. However, with the specified parameter values, the quantum correlation between the cavity and the mechanical mode is less effective, resulting in weak entanglement on the order of $ 10^{-3} $, as shown in Fig. \ref{fig:cavEnM1Gc}(a).
By increasing the Coulomb coupling to $ G_c = 0.5\omega_1 $, the magnitude of entanglement also rises, as evidenced by the broader contour shown in Fig. \ref{fig:cavEnM1Gc}(b). When the Coulomb coupling is further elevated to higher values, specifically $ G_c = 0.6\omega_1 $, $ 0.7\omega_1 $, $ 0.75\omega_1 $, and $ G_c = 0.86\omega_1 $, we observe a corresponding increase in the quantity of entanglement. Additionally, the dark-red region of the contour, representing maximal intensity, gradually shifts to a different location with the varying values of $ G_c $, as illustrated in Figs. \ref{fig:cavEnM1Gc}(c) through (f).
This observation indicates that tuning the Coulomb coupling $ G_c $ from lower to higher values enhances the entanglement and shifts its location, suggesting increased tunability and controllability of our hybrid COMM system.

Next, we examine the robustness of the cavity-mirror entanglement in relation to the magneto-optical coupling values, as illustrated in {Fig. \ref{fig:cavEnM1gm}}. A weak magneto-optical interaction, specifically $ g_m/2\pi = 11 $ MHz, yields cavity-mirror entanglement with a low magnitude, as shown in Fig. \ref{fig:cavEnM1gm}(a). This indicates a relatively weaker electromagnetic drive or magnetic field applied to the COMM, resulting in diminished correlations between photons and phonons. Conversely, with a stronger interaction between photons and magnons, we observe a significant enhancement in correlation, leading to entanglement with a larger magnitude, as depicted in Fig. \ref{fig:cavEnM1gm}(b). Our proposed setup facilitates an indirect interaction of cavity photons with mechanical resonators by modulating the radiation pressure force, a mechanism that is rarely explored in the literature, particularly regarding its role in generating and enhancing entanglement.
\subsection{Influence of Magnon Kerr Nonlinearity $K_s$ on Entanglement Between Cavity Photon and Motional Mode  M$_1$}
\label{results:cavity and M1 vs kerr}
The magnon self-Kerr effect, which induces a frequency shift in the magnon mode, is widely recognized and cannot be overlooked in most scenarios. This shift arises from Kerr nonlinearity, which introduces a self-interaction term in the magnon frequency; consequently, the frequency deviation is proportional to the magnon population $ M $. This nonlinear shift allows for precise control over the magnon’s detuning, a crucial aspect for tuning resonance conditions in hybrid systems and facilitating interactions with other modes, such as photons in our proposed hybrid COMM system.
 \begin{figure*}[t!]
  \includegraphics[width=0.8\linewidth]{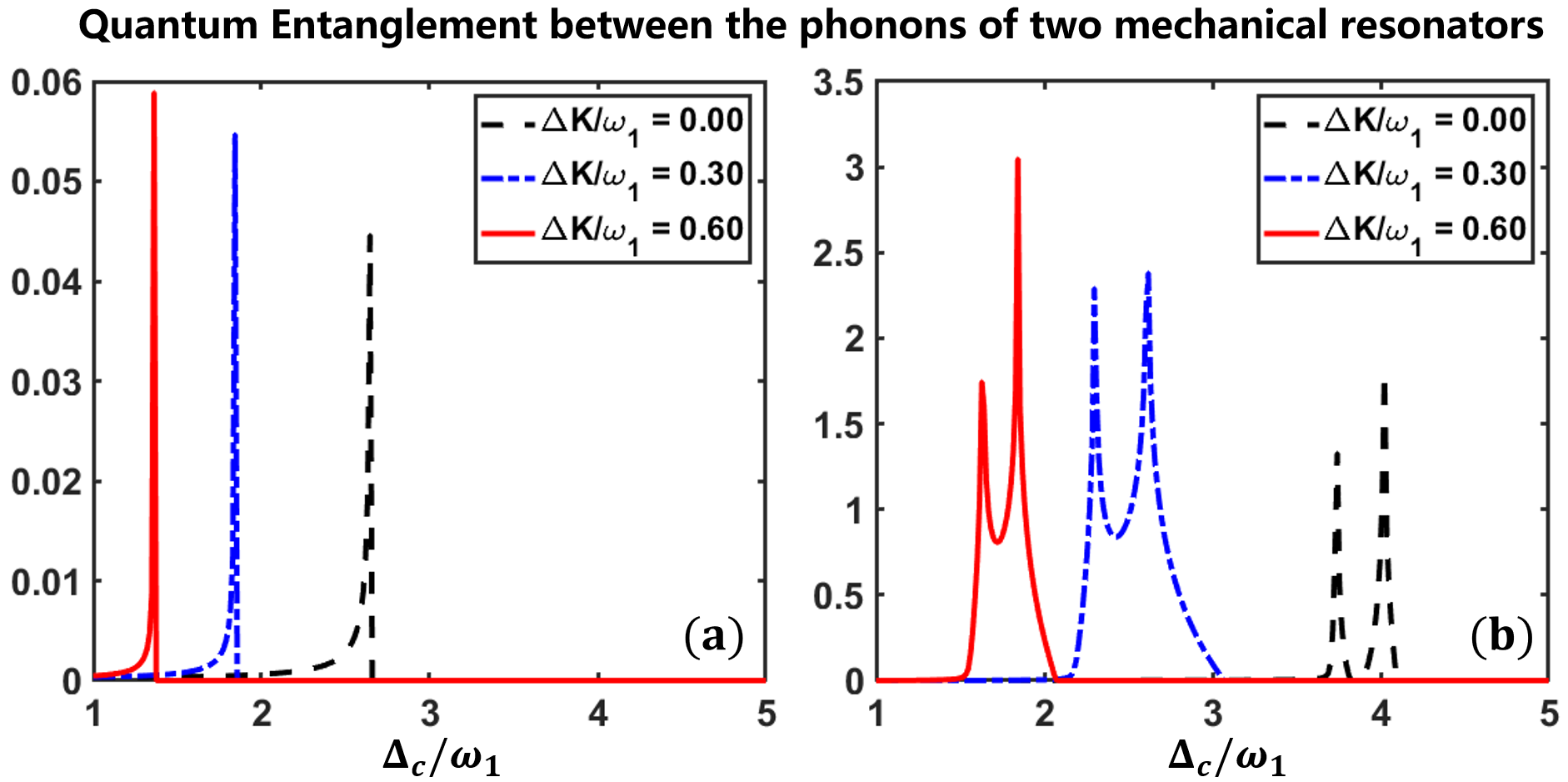}
  \centering
  \caption{Quantity of photon-phonon entanglement plotted against scaled cavity detuning $\Delta_c$, (a) under various values of $\Delta K$ when $G_{\rm eff} = 0.06 \kappa$ (b) under different values of $\Delta K$ when the optomechanical coupling strength is changed to $G_{\rm eff} = 0.25 \kappa$. The value of magnetic dipole interaction used here is $g_m /2\pi = 10$ MHz, and Coulomb coupling $G_c = 0.3 \omega_1$. Other parameters have the same values as given in Fig. \ref{GcControlledEN}.}
  \label{fig:cavEnM1kerr}
\end{figure*}
Our setup comprises three distinct modes: photons, magnons, and phonons. The cavity photon mode is directly coupled to magnons via the magneto-optical interaction $ g_m $ and to phonons through the radiation pressure force, represented by the effective optomechanical coupling $ G_{\rm eff} $. In this study, we concentrate on exploring the entanglement dynamics between the cavity photon and the motional phonon modes of the mirror, aiming to demonstrate how the magnon Kerr nonlinearity $ K_s $ influences this entanglement. Additionally, we emphasize the significance of optomechanical coupling strength in our examination of entanglement.
We adjust the magnon Kerr nonlinearity $ K_s $ in terms of the magnon frequency shift $ \Delta K = 2K_s M $, where this shift modifies the resonance frequency of the magnons, effectively creating a self-induced detuning that is dependent on the magnon population $ M $. This detuning alters the system's response to external drives and its interactions with other components, such as photons in cavity-magnon coupling systems.

Figure \ref{fig:cavEnM1kerr} illustrates the behavior of quantum entanglement between cavity photons and the phonons of mechanical resonator M$_1$ for various values of the magnon frequency shift parameter $ \Delta K $.
When the magnon Kerr nonlinearity is absent in our proposed setup, i.e., $ \Delta K / \omega_1 = 0 $, the photon-phonon entanglement exhibits an extremely small magnitude, as indicated by the black-dashed curve in Fig. \ref{fig:cavEnM1kerr}(a). As the Kerr parameter is increased to $ \Delta K = 0.6 \omega_1 $, the quantity of entanglement is enhanced, and its peak shifts to the left along the cavity detuning spectrum. 
This clearly demonstrates the significant impact of the magnon Kerr effect on the cavity dynamics, as evidenced by the shifting of the entanglement spectrum along the cavity detuning axis. It is important to note that, in this case, the optomechanical coupling strength is relatively low, specifically $ G_{\rm eff} = 0.06 \kappa $, which results in a weak magnitude of entanglement between the photon and phonon modes. To further investigate how the optomechanical coupling influences entanglement enhancement with varying magnon Kerr effects, as previously illustrated in Fig. \ref{fig:cavEnM1kerr}(a), we now increase the optomechanical coupling to $ G_{\rm eff} = 0.25 \kappa $.

Substantial optomechanical coupling enhances the correlation between the cavity photons and the phonons of the mechanical resonator, resulting in significantly pronounced entanglement, as evidenced by its magnitude and the presence of two peaks in Fig. \ref{fig:cavEnM1kerr}(b). In the absence of magnon Kerr nonlinearity, the entanglement curve lies far from the origin along the cavity detuning axis. However, with larger Kerr values, the cavity field resonances shift to new frequencies, causing the entanglement curves to be pulled closer to the origin on the frequency spectrum.

The sign and magnitude of $ K_s $ can be controlled by adjusting the direction and strength of the static magnetic field applied to the YIG sphere, as well as indirectly through the external pump field driving the cavity via the beam-splitter-like interaction between the cavity and the magnon mode ($ c m^{\dagger} + c^{\dagger} m $). The magnon Kerr nonlinearity $ K_s $ is crucial for enhancing quantum entanglement, as it introduces a controllable, nonlinear frequency shift in the magnon mode. By effectively tuning the magnon frequency through the two-magnon interaction, $ K_s $ enables precise control over magnon detuning, thereby facilitating resonance conditions that are favorable for stronger quantum correlations.
\section*{Experimental Feasibility}
Although there is no direct experimental scheme or setup available for the proposed study yet, its feasibility arises from the integration of subsystems that have been validated in previous studies. The coupling between microwave cavities and mechanical resonators in this model is based on the established framework of electro-optomechanical systems, which have been realized in both silicon and superconducting platforms \cite{teufel2011sideband,cattiaux2020beyond,zhao2025quantum,bozkurt2025mechanical}. Additionally, superconducting electromechanical circuits have demonstrated reconfigurable nonreciprocal transmission without the need for magnetic fields, utilizing optomechanical interactions and mechanical mode interference \cite{bernier2017nonreciprocal}.
The magnon-microwave coupling in our system builds on the established interaction between magnons in the YIG sphere and microwave cavity photons, i.e., magnetic-dipole interaction. Previous studies have shown strong coupling mediated by the magnetic dipole interaction between microwave photons and magnons in YIG \cite{zhang2014strongly}. The coupling strength $g_m$ in our model aligns with experimentally achieved rates in similar systems, ensuring its feasibility.
The Coulomb coupling configuration is informed by studies on piezoelectric materials and graphene-integrated electromechanical systems \cite{verbiest2021tunable}. In these systems, electrostatic potentials regulate the coupling strength between mechanical resonators. A similar approach is employed in this setup, where a DC bias tunes the Coulomb interaction strength $G_c$. The application of DC-biased capacitors for this purpose has been validated in MEMS systems and is directly transferable to the present model \cite{zhao2025quantum}.
Operation at cryogenic temperatures ($T = 10~\mathrm{mK}$) reduces thermal noise and facilitates the generation of quantum correlations. The coupling strengths and decay rates ($G_{\mathrm{eff}}$, $G_c$, $\kappa$, $\gamma_m$) can be tuned to match values reported in existing experiments, thereby enabling the observation of entanglement dynamics.

In summary, the proposed system integrates established experimental techniques, providing a practical platform for investigating entanglement in hybrid quantum systems with potential applications in quantum networks and transducers.
\section{Conclusion}
\label{outlook}
In summary, our investigation revealed the generation and enhancement of quantum entanglement between phonon-phonon and photon-phonon modes. To achieve this, we proposed a hybrid Coulomb-enabled cavity-magnon optomechanical system, consisting of an optomechanical cavity enclosing a YIG sphere, where photons and magnons interact via magnetic-dipole coupling. 
An additional mechanical resonator was coupled to the first mechanical resonator via electrostatic coupling $G_c$. Our approach involved coupling the cavity photons with the mechanical resonator M$_1$ and the magnon of the YIG sphere to induce strong correlations among the various modes. We focused on examining the entanglement dynamics across these coupled modes, specifically between the cavity photon and mechanical resonator M$_1$.
Additionally, we discussed in detail the electrostatic interaction between the two charged motional modes and its significance in generating quantum entanglement. The primary goals of our investigation were to examine how various parameters, such as the magnon Kerr nonlinearity, magnetic-dipole coupling, optomechanical coupling strength, and Coulomb interaction, influence the creation and robustness of entanglement within our complex COMM setup.
Our results revealed that the magnon Kerr effect played a pivotal role in adjusting the frequency detuning, which was crucial in modulating both the strength and stability of entanglement. Additionally, the introduction of Coulomb forces between the mechanical resonators enhanced entanglement stability, particularly at elevated bath temperatures, thereby reinforcing the system’s resilience to environmental noise.
Overall, our findings open new avenues for constructing entangled states between distinctly separated resonators, potentially facilitating continuous-variable quantum information tasks, such as quantum memories. This study contributes to the broader understanding of entanglement control in hybrid quantum systems, thereby advancing applications in quantum information science.
\medskip
\section*{\textbf{Conflict of interest}}
The authors declare no financial/commercial conflict of interest.
\appendix
\section{Derivation of the Magnon Driving Hamiltonian}
The physical origin of the coherent magnon driving term is the \textit{Zeeman interaction} between the collective spin angular momentum of the yttrium iron garnet (YIG) sphere and the applied time-dependent microwave magnetic field \cite{li2018magnon}. This microwave field coherently drives the magnon mode. The derivation of the Hamiltonian term for this driving is as follows:

    We start with the Hamiltonian for a spin in a magnetic field, $H = -\gamma\vec{s}\cdot\vec{B}$. For the collective spin angular momentum $\vec{S} = \Sigma \vec{s}$ of the YIG sphere in a drive magnetic field $\vec{B}$ along the $y$-direction, the Hamiltonian is given by:
    \begin{equation}
        H_{d} = -\gamma\vec{S}\cdot\vec{B} = -\gamma S_{y}B_{0}\cos\omega_{B}t,
    \end{equation}
    where $B_0$ is the field amplitude and $\omega_B$ is the frequency of the magnetic field.
    We rewrite the collective spin operator $S_y$ in terms of the raising and lowering operators, $S^{\pm} = S_x \pm iS_y$, which gives $S_y = (S^{+} - S^{-})/2i$. The Hamiltonian then becomes:
    \begin{equation}
        H_{d} = i\frac{\gamma B_{0}}{4}(S^{+} - S^{-})(e^{i\omega_{B}t} + e^{-i\omega_{B}t})
    \end{equation}
    Next, we utilize the \textit{Holstein-Primakoff transformation} to relate the collective spin operators to the bosonic creation ($m^{\dagger}$) and annihilation ($m$) operators of the magnon mode \cite{holstein1940field}. For the low-lying excitations, we make the approximation that $\langle m^\dagger m \rangle \ll 2Ns$, where $N$ is the total number of spins and $s=5/2$ for the $Fe^{3+}$ ion in YIG. This approximation leads to the relations $S^{+} \approx \hbar\sqrt{5N}m$ and $S^{-} \approx \hbar\sqrt{5N}m^{\dagger}$.
     Substituting these relations into the Hamiltonian, we get:
    \begin{equation}
        H_{d}/\hbar = i\frac{\sqrt{5}}{4}\gamma\sqrt{N}B_{0}(m - m^{\dagger})(e^{i\omega_{B}t} + e^{-i\omega_{B}t}).
    \end{equation}
    Finally, by applying the Rotating-Wave Approximation, we only retain the resonant terms that describe the absorption and creation of magnons, leading to the simplified Hamiltonian term:
    \begin{equation}
        H_{d}/\hbar \approx i\Omega_B ( m^{\dagger} e^{-i\omega_{B}t} -m e^{i\omega_{B}t})
    \end{equation}
    where the Rabi frequency $\Omega_B$ is the coupling strength defined as:
    \begin{equation}
    \Omega_B = \frac{\sqrt{5}}{4}\gamma\sqrt{N}B_{0},
    \end{equation}
where $\gamma/2\pi = 28$ GHz/T, $N = \rho V$ with $\rho = 4.22 ~\times 10^{27}$ m$^{-3}$ the YIG's spin density and $V$ the volume of the sphere.

This derivation demonstrates that the driving term stems from the Zeeman interaction, and the underlying physical mechanism is the coherent excitation of the magnon mode by the external microwave magnetic field. The specific form of the Hamiltonian term arises from the Holstein-Primakoff transformation and the Rotating-Wave Approximation.

\section{Frame Rotation}
To eliminate the time dependence from the Hamiltonian in Eq. \eqref{hamiltonian}, we perform a frame rotation with respect to the magnetic drive (laser) field of frequency $\omega_B$ using the unitary operator $U = \text{exp}[-i\omega_B(c^{\dagger} c + m^{\dagger}m)t]$. The resulting frame-rotated Hamiltonian in the interaction picture is:
\begin{equation}
\label{interaction picture}
H_{T}=U(H_{s}-i\hbar\partial_{t})U^{\dagger}
\end{equation}
Next, we substitute the system Hamiltonian from Eq. \eqref{hamiltonian} into the interaction picture transformation given by Eq. \eqref{interaction picture} and evaluate each term individually. For simplicity, we set $\hbar = 1$ throughout the following calculations.
\begin{align}
    \label{interaction picture1}
H_{T} = & U [ \omega_{c} c^{\dagger} c +\omega_{m} m^{\dagger} m + \frac{1}{2}\sum_{j=1}^{2} \omega_j (x_{j}^2 + p_{j}^2) - G_0 c^{\dagger} c x_1 \nonumber\\[4pt]& + G_c x_1 x_2   + K_s (m^{\dagger} m)^2  - g_{m} \left(c^{\dagger} m + m^{\dagger} c \right) \nonumber\\[4pt]& + i \Omega_B \left(m^{\dagger} e^{- i \omega_B t} - m  e^{ i \omega_B t} \right)  - i \hbar \partial_{t}] U^\dagger.
\end{align}
We consider the first term of the Hamiltonian \eqref{interaction picture1}, and use the Baker-Campbell-Hausdorff (BCH) identity \cite{sakurai2020modern}, i.e.,
\begin{align}
\label{baker}
    e^{A} B e^{-A} = B + [A,~B] + \frac{1}{2}[A,[A,~B]] +...,
\end{align}
with $A = {-i\omega_B(c^{\dagger} c + m^{\dagger} m) t}$,  and $B = \omega_{c}c^{\dagger}c$.
By applying the BCH formula in Eq. \eqref{baker} to the first term of Eq. \eqref{interaction picture1}, we get the expanded form as given below.
\begin{align}
\label{B2}
    e^{A} B e^{-A} = & \omega_c c^{\dagger} c + [-i\omega_B(c^{\dagger} c + m^{\dagger} m) t,~\omega_c c^{\dagger} c]  + \frac{1}{2}[ -i\omega_B \nonumber\\[4pt]& (c^{\dagger} c + m^{\dagger} m) t,[ -i\omega_B(c^{\dagger} c + m^{\dagger} m) t,  ~\omega_c c^{\dagger} c]] +....
\end{align}
From the commutation relations of the operators, we know that \cite{susskind2014quantum}
\begin{align}
    &[A,~A] = [A,~B] = 0,~ [A,~A^{\dagger}] = 1, ~ \text{and} ~ [A^{\dagger},~A] = -1.
\end{align}
Using these commutation relations, the whole Equation \eqref{B2} reduces to the expression given below:
\begin{align}
\label{B3}
    e^{-i\omega_B(c^{\dagger} c + m^{\dagger} m) t} (\omega_c c^{\dagger} c) e^{i\omega_B(c^{\dagger} c + m^{\dagger} m) t} = \omega_c c^{\dagger} c.
\end{align}
\noindent The second term of Hamiltonian \eqref{hamiltonian} is calculated in the same way by using the BCH identity, as these operators belong to the same mode (magnon). This would result in the following.
\begin{align}
\label{C1}
    e^{-i\omega_B(c^{\dagger} c + m^{\dagger} m) t} (\omega_m m^{\dagger} m) e^{i\omega_B(c^{\dagger} c + m^{\dagger} m) t} = \omega_m m^{\dagger} m.
\end{align}

\noindent The third expression can be simplified using the BCH formula:
\begin{align}\label{mechanical}
    e^{-i\omega_B(c^{\dagger} c + m^{\dagger} m) t} \left(\frac{1}{2}\sum_{j=1}^{2} \omega_j (x_{j}^2 + p_{j}^2)\right) e^{i\omega_B(c^{\dagger} c + m^{\dagger} m) t}
\end{align}
To solve this, we have the exponent operators $\hat{A} = -i\omega_B(c^{\dagger} c + m^{\dagger} m) t$, and $\hat{B} = \frac{1}{2}\sum_{j=1}^{2} \omega_j (x_{j}^2 + p_{j}^2)$.
The transformation here is of the form $e^{\hat{A}}\hat{B}e^{-\hat{A}}$. Let's evaluate the first commutator, $[\hat{A}, \hat{B}]$.
\begin{align}
    [\hat{A}, \hat{B}] = \left[ i\omega_B(c^{\dagger} c + m^{\dagger} m) t, \frac{1}{2}\sum_{j=1}^{2} \omega_j (x_{j}^2 + p_{j}^2) \right] = 0
\end{align}
Since the commutators approach zero, all higher-order commutators in the BCH expansion will also approach zero. As a result, Eq. \eqref{mechanical} simplifies to just the original operator $\hat{B}$.
\begin{align}\label{mechanical1}
 e^{-i\omega_B(c^{\dagger} c + m^{\dagger} m) t}  \left(\frac{1}{2}\sum_{j=1}^{2} \omega_j (x_{j}^2 + p_{j}^2)\right) e^{i\omega_B(c^{\dagger} c + m^{\dagger} m) t} \nonumber\\  = \frac{1}{2}\sum_{j=1}^{2} \omega_j (x_{j}^2 + p_{j}^2)  
\end{align}

\noindent The optomechanical and Coulomb coupling terms are unaffected by the transformation. Likewise, the term associated with the magnon Kerr coefficient also remains unaltered.

Next, we apply a frame rotation to the photon-magnon coupling term $( g_{m}( c^{\dagger} m + m^{\dagger} c )$. The transformed expression is given by:
\begin{align}
    U [ g_{m}( c^{\dagger} m + m^{\dagger} c ) ] U^\dagger = & g_{m} [ (c^{\dagger} e^{-i\omega_B t})(m e^{i\omega_B t}) \nonumber\\& + (m^{\dagger} e^{-i\omega_B t})(c e^{i\omega_B t}) ]
\end{align}
Since the exponential terms cancel each other out, as $e^{-i\omega_B t}e^{i\omega_B t} = 1$, we obtain the photon-magnon interaction expression unchanged.
\begin{align}
    \label{beam splitter}
  U [ g_{m}( c^{\dagger} m + m^{\dagger} c ) ] U^\dagger  = g_{m} ( c^{\dagger} m + m^{\dagger} c )
\end{align}
The unitary transformation of the magnetic drive laser field and the magnon mode interaction term, i.e., $ i \Omega_B  (m^{\dagger} e^{ i \omega_B t} - m e^{-i\omega_B t})$ removes the time dependency. The derivation is given as
\begin{align}
U\!\left[ i\Omega_B\!\left(m^{\dagger} e^{- i \omega_B t} - m e^{ i \omega_B t} \right) \right] U^\dagger
&= i \Omega_B \Big[ (U m^{\dagger} U^\dagger) e^{- i \omega_B t} \notag\\
&\quad - (U m U^\dagger) e^{ i \omega_B t} \Big].
\end{align}
After simplification, the final expression is given below:
\begin{align}
    \label{drive}
    U \left[ i \Omega_B \left(m^{\dagger} e^{- i \omega_B t} - m e^{ i \omega_B t} \right) \right] U^\dagger = i \Omega_B  (m^{\dagger}  - m ).
\end{align}

Finally, we simplify the partial derivative term in Eq. \eqref{interaction picture}, i.e., $-U \partial_t U^{\dagger}$.
Since the Hermitian conjugate of the unitary operator $U$ is given as:
$$U^{\dagger} = e^{i\omega_B(c^{\dagger}c + m^{\dagger}m)t},$$ we calculate the partial derivative of $U^{\dagger}$ with respect to time $t$:
\begin{align}
    \label{partial}
    \frac{\partial}{\partial t} \left[ e^{i\omega_B(c^{\dagger}c + m^{\dagger}m)t} \right] = i\omega_B(c^{\dagger}c + m^{\dagger}m)U^{\dagger}
\end{align}
The number operators $c^{\dagger}c$ and $m^{\dagger}m$ are invariant under this unitary transformation (since they commute with the unitary operator). Thus, $U (c^{\dagger}c + m^{\dagger}m)U^{\dagger} = i \omega_B (c^{\dagger}c + m^{\dagger}m)$. We substitute this result into the expression for the transformed term:
\begin{align}
\label{partial1}
-i U \frac{\partial U^{\dagger}}{\partial t} = i U \left[ i\omega_B(c^{\dagger}c + m^{\dagger}m)U^{\dagger} \right] = -\omega_B(c^{\dagger}c + m^{\dagger}m).
\end{align}
By adding all these transformed equations, we obtain the frame-rotated, time-independent Hamiltonian given in Eq. \eqref{hamiltonian1} of the main text.
\medskip
\bibliographystyle{apsrev4-2}
\providecommand{\noopsort}[1]{}\providecommand{\singleletter}[1]{#1}%

\end{document}